\documentclass[preprint,review,3p,numbers,sort&compress]{elsarticle}
\usepackage{xcolor}
\usepackage{hyperref}
\usepackage{amsmath,amssymb,amsfonts,amstext,graphicx}
\usepackage{epstopdf}
\usepackage{enumitem}
\usepackage{psfrag}
\usepackage{array}
\usepackage{epstopdf}
\usepackage{flushend}
\usepackage{booktabs}
\usepackage{verbatim}
\usepackage{cases}
\usepackage{MnSymbol}
\usepackage{float}
\usepackage{longtable}
\usepackage[center]{caption2}
\allowdisplaybreaks
\newcommand{\vecnm}[1]{\left\|#1\right\|}

\begin{document}

\begin{frontmatter}

\title{MIMO radar target localization and performance evaluation under SIRP clutter}

\author[mymainaddress]{Xin Zhang\corref{mycorrespondingauthor}}
\cortext[mycorrespondingauthor]{Corresponding author}
\ead{xzhang@nt.tu-darmstadt.de}

\author[mysecondaryaddress]{Mohammed Nabil El Korso}

\author[mymainaddress]{Marius~Pesavento}

\address[mymainaddress]{Communication System Group, Technische Universit\"{a}t Darmstadt, Darmstadt, 64283, Germany}
\address[mysecondaryaddress]{Laboratoire Energ\'{e}tique M\'{e}canique Electromagn\'{e}tisme, Universit\'{e} Paris Ouest Nanterre La D\'{e}fense, 92410, Ville d{'}Avray, France}

\begin{abstract}
Multiple-input multiple-output (MIMO) radar has become a thriving subject of research during the past decades. In the MIMO radar context, it is sometimes more accurate to model the radar clutter as a non-Gaussian process, more specifically, by using the spherically invariant random process (SIRP) model. In this paper, we focus on the estimation and performance analysis of the angular spacing between two targets for the MIMO radar under the SIRP clutter. First, we propose an iterative maximum likelihood as well as an iterative maximum a posteriori estimator, for the target's spacing parameter estimation in the SIRP clutter context. Then we derive and compare various Cram\'{e}r-Rao-like bounds (CRLBs) for performance assessment. Finally, we address the problem of target resolvability by using the concept of angular resolution limit (ARL), and derive an analytical, closed-form expression of the ARL based on Smith's criterion, between two closely spaced targets in a MIMO radar context under SIRP clutter. For this aim we also obtain the non-matrix, closed-form expressions for each of the CRLBs. Finally, we provide numerical simulations to assess the performance of the proposed algorithms, the validity of the derived ARL expression, and to reveal the ARL's insightful properties.
\end{abstract}

\begin{keyword}
multiple-input multiple-output radar, spherically invariant random process, maximum likelihood estimation, maximum a posteriori estimation, Cram\'{e}r-Rao-like bounds, angular resolution limit
\end{keyword}

\end{frontmatter}


\section{Introduction}

During the past decade, multiple-input multiple-output (MIMO) radar has been attracting an increasing academic interest \cite{mim1,LS08}. MIMO radar, as opposed to conventional phased-array radar, can exploit multiple antennas both to simultaneously transmit orthogonal waveforms and also to receive the reflected signals. By virtue of this waveform diversity, MIMO radar enables to significantly ameliorate the performance of radar systems, in terms of improved parameter identifiability, more flexible beam-pattern design, direct applicability of space-time adaptive processing techniques, \cite{BT06,LS08,mim2} etc. Abounding works have been dedicated to MIMO radar, either to investigate algorithms for target localization or to evaluate their performances in terms of lower bounds or resolvability \cite{LS08,HBC08,mim3,BT06,JLL09,mim2,zhxtc8,zhxtc6,mim18,xia1,einemo1,hong1,tang1}. In the larger part of the radar literature, the clutter is simply assumed to be a Gaussian stochastic process. Such assumption is generally a good approximation in many cases and has its theoretical basis in the central limit theorem. However, in certain specific scenarios, the radar clutter cannot be correctly described by the Gaussian model anymore. As an example, experimental measurements reveal that the ground clutter data heavily deviate from the Gaussian model \cite{mim4}. This is also true, e.g., for the sea clutter in a high-resolution and low-grazing-angle radar context, where the scatter number is random and the clutter shows nonstationarity \cite{mim5}.

To account for such problems, where the clutter is a non-Gaussian process, numerous clutter models have been developed. Among them, the so-called spherically invariant random process (SIRP) model has become the most notable and popular one in radar clutter modeling \cite{mim4,mim5,mim6,mim11,mim12}. Its main advantage lies in its feasibility to describe different scales of the clutter roughness, as well as its generality to encompass a wide variety of non-Gaussian distributions (K-distribution, t-distribution, Laplace, Cauchy and Weibull distribution, etc.). A SIRP is a two-scale, complex, compound Gaussian process with random power, structured as the product of two components: a complex Gaussian process with zero mean and unknown covariance matrix, and the square root of a positive scalar random process \cite{mim6}. In the radar context, the former describes the local scattering and is usually referred to as \textit{speckle}, while the latter, modeling the local power changing, is called \textit{texture}. A SIRP is fully characterized by its texture parameter(s) and the covariance matrix of its speckle. Till now, the SIRP model has gained widespread use to treat the heavy-tailed, non-Gaussian distributions of radar clutters \cite{mim10,mim7,mim5,mim8,mim9}.

Not few works have addressed the estimation problems associated with the SIRP clutter. Most of them deal solely with the estimation of clutter parameters. Specifically, the texture parameter(s) and/or the speckle covariance are estimated, by assuming the presence of secondary data (known noise-only realizations) in designing their algorithms \cite{mim13,mim14,mim15,mim17,lombardo1,mim8}. However, in our context, we consider unknown clutter realizations embedded
in and contaminating the received signal. Furthermore, we are interested in the target's spacing parameter instead of the unknown clutter nuisance parameters. In \cite{mim15} and \cite{Akcakaya1}, on the other hand, the authors devised parameter-expanded expectation-maximization (PX-EM) algorithms to estimate the signal as well as clutter parameters for the traditional phased-array radar and MIMO radar, respectively. Nevertheless, the algorithms proposed in \cite{mim15} and \cite{Akcakaya1} are restricted to a special, linear signal model, called the generalized multivariate analysis of variance (GMANOVA) model \cite{dog1}, under which category our context does not fall. To the best of our knowledge, no available algorithm in the current literature addresses the target estimation problem, or the problem of the direction-of-departure/arrival (DOD/DOA) estimation \cite{HPRE14} (a highly non-linear problem) in general, under the SIRP clutter in a comprehensive manner. In this paper, we devise an iterative maximum likelihood estimator (IMLE), together with an iterative maximum a posteriori estimator (IMAPE), to serve such a purpose. Our algorithms carry on the path trodden by \cite{PG01} and \cite{vorobyov1} and can be seen as generalizations of them, due to their common iterative nature and the idea of \emph{stepwise concentration}. To evaluate the performance of our algorithms, we further derive expressions for the standard Cram\'{e}r-Rao bound (CRB) and for its variants, including the extended Miller-Chang bound (EMLB), the modified CRB (MCRB) and the hybrid CRB (HCRB), w.r.t. the target's spacing parameter. We then provide an extended examination of their relationships, and the relationships between them and the texture parameters.

Furthermore, in order to fully characterize the performance analysis, we further investigate the resolvability problem of two closely spaced targets. In the MIMO radar context, a few recent works, e.g., \cite{zhxtc8} and \cite{zhxtc6,mim18}, have addressed this problem. The clutter in these works, however, is unexceptionally modeled as a Gaussian process. In this paper, we take on the resolvability problem concerning two (colocated) MIMO radar targets under non-Gaussian clutter (modeled as SIRP). To be more specific, this paper sets as its principal aim the solution to the following question: ``\textit{What is, in a colocated MIMO radar context under non-Gaussian clutter, the minimum angular separation (between two closely spaced targets) required, under which these two targets can still be correctly resolved?}" No work in the current literature, to the best of our knowledge, has been dedicated to this question, except our preliminary work \cite{xzh3}, in which we approached this problem by numerical means. In this paper, we carry on with what was set out in \cite{xzh3} and bring it to completion, by proposing an analytical expression as the solution to the question under discussion, and by considering a wider range of clutter distributions.

To approach this question we resort, in a similar way to \cite{zhxtc8,zhxtc6,mim18}, to the concept of the resolution limit (RL), which provides the theoretical foothold of our work to characterize the resolvability of two targets. The RL is defined as the minimum distance w.r.t. the parameter of interest (e.g., the DODs/DOAs or the electrical angles, etc.) that allows distinguishing between two closely spaced sources \cite{S05,ShaMil05,EBRM10}. Various approaches have been devised to account for the RL, generally categorized, in view of the respective theories they rest on, into three families: those based on the mean null spectrum analysis \cite{C73}, those capitalizing on the detection theory \cite{ShaMil05,SM04,LN07,elkorso1,elkorso2}, and finally, those concerning the estimation theory and exploiting the CRB \cite{L92,S05,EBRM11a,xzh1,xzh2}. Belonging to the family of the third approach, a widely recognized criterion is proposed by Smith \cite{S05}, according to which two targets {\it are resolvable if the distance between the targets (w.r.t. the parameter of interest) is greater than the standard deviation of the distance estimation}. The prevalence of Smith's criterion, over other criteria derived from the estimation theory, e.g., the one proposed in \cite{L92,L94,D98}, is largely attributable to its merit of taking the coupling between the parameters into account. Moreover, it enjoys generality in contrast to the mean null spectrum approach, as the latter is designed for certain specific high-resolution algorithms and not for a specific signal model itself \cite{zhxtc9}. Finally, the RL yielded by Smith's criterion is closely related, as recently revealed in \cite{EBRM10}, to the class of the detection theory based approach, meaning that these two approaches can in fact be unified. In view of these merits, we focus on the RL in Smith's sense in this paper. First, we propose an analytical expression for the angular resolution limit (ARL\footnote{The so-called ARL refers to the RL when angular parameters are considered as the only parameters of interest.}) between two closely-spaced targets in a colocated MIMO radar system under SIRP clutter. As a byproduct, closed-form expressions of the standard CRB w.r.t. the angular spacing are derived. Furthermore, we provide numerical illustrations to vindicate our expression, as well as to inspect the properties revealed by it.

The remaining part of this paper is structured as follows. Section \ref{sec2} introduces the observation model of the colocated MIMO radar system and specifies the observation statistics. In Section \ref{sec3} and Section \ref{sec3b}, our proposed IMLE and IMAPE are respectively derived. Section \ref{sec4} presents the expressions of the Cram\'{e}r-Rao-like bounds (CRLBs) and provides analytical results on their respective properties. Section \ref{sec5} is dedicated to the derivation of analytical expression of the ARL. Section \ref{sec6} provides the simulation results and discusses the properties of our estimator, bounds and the ARL revealed by the figures. Finally, Section \ref{sec_con} summarizes the work of this paper.

\section{Model Setup}\label{sec2}
\subsection{Observation Model for Colocated MIMO Radar}
Consider a colocated MIMO radar system with linear, possibly non-uniform, arrays both at the transmitter and the receiver. Two targets are illuminated by the MIMO radar, both modeled as far-field, narrowband, point sources \cite{LS08}. Furthermore, consider, for simplicity of description, that there is one radar pulse in a coherent processing interval (CPI)\footnote{Note that our derivations and results in this paper can be generalized to the case where more than one pulse per CPI is considered.}. The radar output, without matched filtering, is given as the following vector form \cite{mim2}:

\begin{equation}\label{1b}
\boldsymbol{y}(t)=\sum_{i=1}^{2}\alpha_{i}\boldsymbol{a}_{\mathcal R}\left(\omega_{i}\right)\boldsymbol{a}_{\mathcal T}^{T}\left(\omega_{i}\right)\boldsymbol{s}(t)+\boldsymbol{n}(t),\quad t=1,\dots,T.
\end{equation}
where $\alpha_{i}$ and $\omega_{i}$ denote a complex coefficient proportional to the radar cross section (RCS) and the electrical angle\footnote{Note that, since we are considering a colocated MIMO radar, a target has the same electrical angle at the transmitter and the receiver.} of the $i$th target, respectively; $T$ denotes the number of snapshots per pulse; the transmit and receive steering vectors are defined as
$\boldsymbol{a}_{\mathcal T}(\omega_{i})=[e^{j\omega_{i}d_{1}^{(\mathcal T)}},\dots, e^{j\omega_{i}d_{M}^{(\mathcal T)}}]^{T}$ and $\boldsymbol{a}_{\mathcal R}(\omega_{i})=[e^{j\omega_{i}d_{1}^{(\mathcal R)}},\dots, e^{j\omega_{i}d_{N}^{(\mathcal R)}}]^{T}$, in which $M$ and $N$ represent the number of sensors at the transmitter and the receiver, respectively; $d_{i}^{(\mathcal T)}$ and $d_{i}^{(\mathcal R)}$ denote the distance between the $i$th sensor and the reference sensor, for the transmitter and the receiver, respectively; $\boldsymbol{s} (t)=\left[s_{1}(t),\dots, s_{M}(t)\right]^{T}$ and $\boldsymbol{n} (t), \ t=1,\dots,T$ denote the signal target source vectors and the received clutter vectors, respectively; and $(\cdot)^{T}$ denotes the transpose of a matrix.

\subsection{Observation Statistics}\label{IIB}
The signal target source vectors $\boldsymbol{s}(t),\ t=1,\dots,T$ are viewed as deterministic, while the received clutter vectors $\boldsymbol{n}(t),\ t=1,\dots,T$ are assumed to be independent, identically distributed (i.i.d.) spherically invariant random vectors (SIRVs) \cite{mim6}, modeled as the product of two components statistically independent of each other:
\begin{equation}\label{5}
\boldsymbol{n}(t)=\sqrt{\tau(t)}\boldsymbol{x}(t),\quad t=1,\dots,T;
\end{equation}where the texture terms $\tau(t),\ t=1,\dots,T$, are i.i.d. positive random variables, and the speckle terms $\boldsymbol{x}(t),\ t=1,\dots,T$, are i.i.d. $N$-dimensional circular complex Gaussian vectors with zero mean and second-order moments:
\begin{equation}\label{5b}
\begin{aligned}
&\text{E}\left\{\boldsymbol{x}(i)\boldsymbol{x}^H(j)\right\}=\check{\delta}_{ij}\boldsymbol{\Sigma}
=\check{\delta}_{ij}\sigma^2\check{\boldsymbol{\Sigma}},\\ &\text{E}\left\{\boldsymbol{x}(i)\boldsymbol{x}^T(j)\right\}=\boldsymbol{0}_{N\times N}, \quad i,j=1,\dots,T;
\end{aligned}
\end{equation}
in which $\boldsymbol{\Sigma}$ denotes the speckle covariance matrix, $\text{E}{\{\cdot\}}$ is the expectation operator, $(\cdot)^{H}$ denotes the conjugate transpose of a matrix, $\check{\delta}_{ij}$ is the Kronecker delta, $\sigma^2$ is a scale factor to adjust the clutter power, $\check{\boldsymbol{\Sigma}}$ is the normalized $\boldsymbol{\Sigma}$ with $\text{tr}\{\check{\boldsymbol{\Sigma}}\}=1$, where $\text{tr}\{\cdot\}$ represents the trace of a matrix, and $\boldsymbol{0}_{N\times N}$ denotes the $N\times N$ zero matrix.

In this paper, we mainly focus on two kinds of SIRP clutters, namely, the K-distributed and the t-distributed clutters. In both cases the texture is characterized by two parameters, the \emph{shape parameter} $a$ and the \emph{scale parameter} $b$. Thus, the texture pdf is denoted by $p_{\tau(t)}(\tau(t);a,b)$:
\begin{itemize}
\item \textbf{K-distributed clutter}, in which $\tau(t)$ follows the \emph{gamma distribution}, i.e., $\tau(t)\sim \text{Gamma}(a, b)$ \cite{watts1,nohara1,mim5,sangston1}, namely,
\begin{equation}
\begin{aligned}
p_{\tau(t)}(\tau(t);a,b)&=\frac{1}{\Gamma(a)b^a}\tau(t)^{a-1}e^{-\frac{\tau(t)}{b}},
\end{aligned}
\end{equation}
in which $\Gamma(a)=\int_{0}^{+\infty}x^{a-1}e^{-x}\text{d}x$ denotes the gamma function.
\item \textbf{t-distributed clutter}, in which $\tau(t)$ follows the \emph{inverse-gamma distribution}\footnote{Equivalently, $1/\tau(t)$ follows a gamma distribution.}, i.e., $\tau(t)\sim \text{Inv-Gamma}(a, b)$ \cite{mim13,lange1,jay1,jay2}, thus,
\begin{equation}
\begin{aligned}
p_{\tau(t)}(\tau(t);a,b)&=\frac{b^a}{\Gamma(a)}\tau(t)^{-a-1}e^{-\frac{b}{\tau(t)}}.
\end{aligned}
\end{equation}
\end{itemize}

\subsection{Unknown Parameter Vector}
Assume, in the above model, both the target amplitudes $\alpha_{1}$ and $\alpha_{2}$ to be arbitrary, deterministic, unknown complex parameters. We consider the electric angle $\omega_{1}$ to be known while $\omega_{2}$ is unknown\footnote{This assumption makes good sense in many scenarios, e.g., in those where $\omega_{1}$ is considered a friend target whose position is known and $\omega_{2}$ represents the unknown position of the enemy.}. Furthermore, for the convenience of later derivation, let $\Delta=\omega_{2}-\omega_{1}$ denote the angular spacing between the two targets. Consequently, Eq.~(\ref{1b}) becomes:

\begin{equation}\label{1c}
\boldsymbol{y}(t)=\boldsymbol{v}(t)+\boldsymbol{n}(t),\quad t=1,\dots,T;
\end{equation}in which $\boldsymbol{v}(t)=\alpha_{1} \boldsymbol{a}_{\mathcal R}\left(\omega_{1}\right)\boldsymbol{a}_{\mathcal T}^{T}\left(\omega_{1}\right)\boldsymbol{s}(t)+\alpha_{2} \boldsymbol{a}_{\mathcal R}\left(\omega_{1}+\Delta\right)\allowbreak\cdot\boldsymbol{a}_{\mathcal T}^{T}\left(\omega_{1}+\Delta\right)\boldsymbol{s}(t)$ denotes the target component in the observation. Let us introduce a vector parameter $\boldsymbol{\mu}=\left[\Delta, \quad \overline{\alpha}_1, \quad \widetilde{\alpha}_1, \quad \overline{\alpha}_2, \quad \widetilde{\alpha}_2\right]^T$ which contains all the unknown real target parameters, in which $\overline{(\cdot)}$ and $\widetilde{(\cdot)}$ represent the real and the imaginary part, respectively.

With regard to the SIRP clutter, assume both of its texture parameters, $a$ and $b$, as well as its speckle covariance matrix $\boldsymbol{\Sigma}$, to be unknown. In addition, we introduce the $N^2$-element vector parameter $\boldsymbol{\zeta}$ containing the real and imaginary parts of the entries of the lower triangular part of $\boldsymbol{\Sigma}$. Consequently, the full unknown parameter vector of our problem is given by:

\begin{equation}
\boldsymbol{\xi}=\left[\boldsymbol{\mu}^T,\boldsymbol{\zeta}^T,a,b\right]^T,
\end{equation}in which $\Delta$ is our parameter of interest.

\subsection{Likelihood Functions}
Let $\boldsymbol y=\left[\boldsymbol{y}^T(1),...,\boldsymbol{y}^T(T)\right]^T$ denote the full observation vector, and $\boldsymbol{\tau}=\left[\tau(1),\dots,\tau(T)\right]^T$ represent the texture vector containing the texture components from all snapshots. Since the clutter vectors of different snapshots are i.i.d., the full observation likelihood conditioned on $\boldsymbol{\tau}$ is:
\begin{equation}\label{n1}
p_{\boldsymbol y| \boldsymbol{\tau}}\left(\boldsymbol y | \boldsymbol{\tau}; \boldsymbol{\psi} \right)=\prod_{t=1}^T\frac{\exp\left(-\frac{1}{\tau(t)}\boldsymbol{\beta}^H(t)\boldsymbol{\beta}(t)\right)}
{\mid\pi\tau(t)\boldsymbol{\Sigma}\mid};
\end{equation}in which $\boldsymbol{\psi}=\left[\boldsymbol{\mu}^T,\boldsymbol{\zeta}^T\right]^T$, and $\boldsymbol{\beta}(t)=
\boldsymbol{\Sigma}^{-1/2}\left(\boldsymbol{y}(t)-\boldsymbol{v}(t)\right)$, standing for the clutter spatially whitened by its speckle covariance matrix, at snapshot $t$.

Multiplying $p_{\boldsymbol y| \boldsymbol{\tau}}\left(\boldsymbol y | \boldsymbol{\tau}; \boldsymbol{\psi} \right)$ by $p_{\boldsymbol \tau}(\boldsymbol \tau;a,b)$ (which is equal to $\prod_{t=1}^T p_{\tau(t)}(\tau(t);a,b)$, as the texture components are i.i.d.) leads to the joint likelihood between $\boldsymbol y$ and $\boldsymbol{\tau}$, viz.:
\begin{equation}\label{n1a}
\begin{aligned}
&p_{\boldsymbol y, \boldsymbol{\tau}}\left(\boldsymbol y,  \boldsymbol{\tau}; \boldsymbol{\xi} \right)
=p_{\boldsymbol y| \boldsymbol{\tau}}\left(\boldsymbol y | \boldsymbol{\tau}; \boldsymbol{\psi} \right)p_{\boldsymbol \tau}(\boldsymbol \tau;a,b)\\
=&\prod_{t=1}^T\frac{\exp\left(-\frac{1}{\tau(t)}
\boldsymbol{\beta}^H(t)\boldsymbol{\beta}(t)\right)}{\mid\pi\tau(t)\boldsymbol{\Sigma}\mid}p_{\tau(t)}(\tau(t);a,b).
\end{aligned}
\end{equation}

Finally, the marginal likelihood, w.r.t. $\boldsymbol{\xi}$, is obtained by integrating out $\boldsymbol{\tau}$ from Eq.~(\ref{n1a}):
\begin{equation}\label{n1b}
\begin{aligned}
&p_{\boldsymbol y}\left(\boldsymbol y; \boldsymbol{\xi} \right)=
\int_0^{+\infty}p_{\boldsymbol y, \boldsymbol{\tau}}\left(\boldsymbol y,  \boldsymbol{\tau}; \boldsymbol{\xi} \right)\text{d}{\boldsymbol \tau}\\
=&\prod_{t=1}^T\frac{\int_{0}^{+\infty}\frac{\exp\left(-\frac{1}{\tau(t)}
\boldsymbol{\beta}^H(t)\boldsymbol{\beta}(t)\right)}{\tau^{N}(t)}
p_{\tau(t)}(\tau(t);a,b)\text{d}\tau(t)}{\mid\pi\boldsymbol{\Sigma}\mid}.
\end{aligned}
\end{equation}

\section{Iterative Maximum Likelihood Estimator}\label{sec3}
To over come the difficulty in maximizing the intractable marginal likelihood function Eq.~(\ref{n1b}), various estimation procedures in the SIRP context have chosen to maximize, instead, either the joint likelihood Eq.~(\ref{n1a}) \cite{mim15}, or the conditional likelihood Eq.~(\ref{n1}) \cite{conte2}. The latter approach treats $\boldsymbol \tau$ as \emph{deterministic}, i.e., one realization from the texture process rather than the process itself. In deriving our IMLE we adopt this idea and the usage of the term \emph{maximum likelihood estimator} (MLE) is with regard to this kind of deterministic texture modeling.

From Eq.~(\ref{n1a}) arises the conditional log-likelihood (LL) function, denoted by $\Lambda_\text{C}$, as:
\begin{equation} \label{ll_con}
\begin{aligned}
\Lambda_\text{C}=&
\ln p_{\boldsymbol y| \boldsymbol{\tau}}\left(\boldsymbol y | \boldsymbol{\tau}; \boldsymbol{\psi} \right)
=-TN\ln\pi-T\ln|\boldsymbol{\Sigma}|\\
&-N\sum_{t=1}^{T}\ln\tau(t)
-\sum_{t=1}^{T}\frac{1}{\tau(t)}\boldsymbol{\beta}^H(t)\boldsymbol{\beta}(t).
\end{aligned}
\end{equation}

Equating $\partial\Lambda_\text{C}/\partial\tau(t)$ to zero leads to $\tau(t)$'s estimate when $\boldsymbol{\mu}$ and $\boldsymbol{\zeta}$ are fixed. This, denoted by $\hat{\tau}(t)$, is given by:
\begin{equation}\label{tau_es}
\begin{aligned}
\hat{\tau}(t)
=\frac{1}{N}\left(\boldsymbol{y}(t)-\boldsymbol{v}(t)\right)^H\boldsymbol{\Sigma}^{-1}
\left(\boldsymbol{y}(t)-\boldsymbol{v}(t)\right).
\end{aligned}
\end{equation}
On the other hand, the estimate of $\boldsymbol{\Sigma}$, denoted by $\hat{\boldsymbol{\Sigma}}$, when $\boldsymbol \mu$ and $\boldsymbol \tau$ and are fixed, can be found by applying Lemma 3.2.2. in \cite{And84} to Eq.~(\ref{ll_con}), as:
\begin{equation}\label{Sig_es}
\hat{\boldsymbol{\Sigma}}=\frac{1}{T}\sum_{t=1}^{T}\frac{1}{\tau(t)}
\left(\boldsymbol{y}(t)-\boldsymbol{v}(t)\right)
\left(\boldsymbol{y}(t)-\boldsymbol{v}(t)\right)^H.
\end{equation}
Plugging Eq.~(\ref{tau_es}) into Eq.~(\ref{Sig_es}), we obtain the following iterative expression of $\hat{\boldsymbol{\Sigma}}$:
\begin{equation}\label{Sig_es2}
\hat{\boldsymbol{\Sigma}}^{(i+1)}=\frac{N}{T}\sum_{t=1}^{T}
\frac{\left(\boldsymbol{y}(t)-\boldsymbol{v}(t)\right)\left(\boldsymbol{y}(t)-\boldsymbol{v}(t)\right)^H}
{\left(\boldsymbol{y}(t)-\boldsymbol{v}(t)\right)^H\left(\hat{\boldsymbol{\Sigma}}^{(i)}\right)^{-1}
\left(\boldsymbol{y}(t)-\boldsymbol{v}(t)\right)},
\end{equation}for which the initialization matrix $\hat{\boldsymbol{\Sigma}}^{(0)}=\boldsymbol{I}_N$, where $\boldsymbol{I}_N$ represents the identity matrix of size $N$.

Iteration (\ref{Sig_es2}) was first derived in \cite{gini1}, and then proved in \cite{mim14} to be the exact maximum likelihood (ML) estimator of $\hat{\boldsymbol{\Sigma}}$ when the vector $\boldsymbol \tau$ is assumed to be deterministic, as is in our current case. The convergence properties of the iteration have been analyzed in \cite{gini1,mim14}.

To make the clutter parameters uniquely identifiable, the scaling ambiguity in the clutter model needs to be resolved. Towards this aim, we stipulate for our estimation problem that $\text{tr}\{\boldsymbol{\Sigma}\}=1$, i.e., $\sigma^2=1$ in Eq.~(\ref{5b}). Thus $\hat{\boldsymbol{\Sigma}}^{(i+1)}$, in Eq.~(\ref{Sig_es2}), needs to be further normalized as:
\begin{equation}\label{Sig_es_n}
\hat{\boldsymbol{\Sigma}}^{(i+1)}_\text{n}=
\frac{\hat{\boldsymbol{\Sigma}}^{(i+1)}}{\text{tr}\left\{\hat{\boldsymbol{\Sigma}}^{(i+1)}\right\}},
\end{equation}in which $\hat{\boldsymbol{\Sigma}}^{(i+1)}_\text{n}$ denotes the normalized $\hat{\boldsymbol{\Sigma}}^{(i+1)}$.

Now, let us consider the estimation of the target parameters $\boldsymbol \mu$. To begin with, we reformulate the expression of $\boldsymbol{v}(t)$ as:
\begin{equation}\label{v}
\boldsymbol{v}(t)=\boldsymbol{B}(t, \Delta)\boldsymbol{\alpha},\quad t=1,\dots,T,
\end{equation}in which $\boldsymbol{\alpha}=\left[\alpha_{1},\alpha_{2}\right]^T$ and $\boldsymbol{B}(t, \Delta)=\left[\boldsymbol{b}_1(t),\boldsymbol{b}_2(t, \Delta)\right]$, where $\boldsymbol{b}_1(t)=\boldsymbol{a}_{\mathcal R}\left(\omega_{1}\right)\boldsymbol{a}_{\mathcal T}^{T}\left(\omega_{1}\right)\boldsymbol{s}(t)$ and $\boldsymbol{b}_2(t, \Delta)=\boldsymbol{a}_{\mathcal R}\left(\omega_{1}+\Delta\right)\allowbreak\cdot\boldsymbol{a}_{\mathcal T}^{T}\left(\omega_{1}+\Delta\right)\boldsymbol{s}(t)$.

The ML estimate of $\boldsymbol{\alpha}$, when $\Delta$, $\boldsymbol \tau$ and $\boldsymbol{\zeta}$ are fixed, is given by the solution of $\partial \Lambda_\text{C}/\partial \boldsymbol{\alpha}=0$ calculated from Eq.~(\ref{ll_con}). We denote this estimate by $\hat{\boldsymbol{\alpha}}$, which has the following expression:
\begin{equation}\label{alpha_es}
\hat{\boldsymbol{\alpha}}=\left(\check{\boldsymbol{B}}^H(\Delta)
\check{\boldsymbol{B}}(\Delta)\right)^{-1}\check{\boldsymbol{B}}^H(\Delta)\check{\boldsymbol{y}},
\end{equation}in which $\check{\boldsymbol{y}}=\boldsymbol{G}^{-1/2}\boldsymbol{y}$, $\check{\boldsymbol{B}}(\Delta)=\boldsymbol{G}^{-1/2}\boldsymbol{B}(\Delta)$, where $\boldsymbol{G}=\text{diag}\{\tau(1),\dots,\tau(T)\}\otimes\boldsymbol{\Sigma}$, in which $\otimes$ denotes the Kronecker product, and $\text{diag}\{\cdot\}$ represents the diagonal matrix whose diagonal entries are arguments inside $\{\cdot\}$. Furthermore, the matrix $\boldsymbol{B}(\Delta)=\left[\boldsymbol{b}_1,\boldsymbol{b}_2(\Delta)\right]$, in which $\boldsymbol{b}_1=\left[\boldsymbol{b}_1^T(1),\dots,\boldsymbol{b}_1^T(T)\right]^T$ and $\boldsymbol{b}_2(\Delta)=\left[\boldsymbol{b}_2^T(1, \Delta),\dots,\boldsymbol{b}_2^T(T, \Delta)\right]^T$. We note that the matrix $\boldsymbol{G}$ serves the purpose of \emph{de-texturizing} and \emph{pre-whitening}.

It is apparent from Eqs.~(\ref{tau_es}), (\ref{Sig_es2}) and (\ref{alpha_es}) that the estimation of the involved parameters are \emph{mutually dependent}, in the sense that the expression for the estimate of any of these parameters contain all the rest of them. In \cite{mim15} and \cite{Akcakaya1}, the authors overcame the similar difficulty by exploiting the special structure of their GMANOVA model and obtained an expression of $\hat{\boldsymbol{\Sigma}}$ independent of their unknown signal parameters. However, such analytical concentration approach is inapplicable to the estimation problem under consideration. Therefore, in this paper we adopt the so-called \emph{stepwise numerical concentration} method, whose concept was introduced and employed, in the context of non-uniform white Gaussian noise in \cite{PG01}, and colored Gaussian noise in \cite{vorobyov1}.

The idea of the stepwise concentration consists in the concentration of the LL function w.r.t. certain unknown parameters in an iterative manner. In our case, we assume for each iteration that $\boldsymbol{\Sigma}$ and $\boldsymbol{\tau}$ are fixed and known, and use their values to compute the estimate of $\boldsymbol{\mu}$, which is then used, in its turn, to update the values of $\boldsymbol{\Sigma}$ and $\boldsymbol{\tau}$ for the next iteration. We continue this procedure until convergence, which can be defined, e.g., by the criterion that the difference between the values of estimates obtained from consecutive iterations fall below a certain small threshold.

This general procedure borne in mind, we return to the LL function in Eq.~(\ref{ll_con}). Now, our aim is to find the estimate of $\Delta$, our parameter of interest, by considering the values of $\boldsymbol{\Sigma}$ and $\boldsymbol{\tau}$ as fixed and known from the previous iteration. Thus, neglecting the constant terms, the conditional LL function in Eq.~(\ref{ll_con}) can be reformulated as:
\begin{equation} \label{ll_con2}
\Lambda_\text{C}=
-\sum_{t=1}^{T}\frac{1}{\tau(t)}\boldsymbol{\beta}^H(t)\boldsymbol{\beta}(t).
\end{equation}
Inserting Eq.~(\ref{alpha_es}) into Eq.~(\ref{ll_con2}) and maximizing the latter w.r.t. $\Delta$ leads to the following estimate:
\begin{equation}\label{Delta_es}
\hat{\Delta}=\arg\min_{\Delta}\left\{\vecnm{\boldsymbol{\Pi}_{\check{\boldsymbol{B}}(\Delta)}^{\bot}
\check{\boldsymbol{y}}}\right\},
\end{equation}in which $\vecnm{\cdot}$ denotes the Euclidean norm and
\begin{equation}
\boldsymbol{\Pi}_{\check{\boldsymbol{B}}(\Delta)}^{\bot}=\boldsymbol{I}_{NT}-\check{\boldsymbol{B}}(\Delta)\left(\check{\boldsymbol{B}}^H(\Delta)
\check{\boldsymbol{B}}(\Delta)\right)^{-1}\check{\boldsymbol{B}}^H(\Delta),
\end{equation}stands for the orthogonal projection matrix onto the null space of $\check{\boldsymbol{B}}(\Delta)$.

Consequently, our proposed IMLE, which consist of three steps, can be summarized as follows:
\begin{itemize}
\item \textbf{Step 1}: Initialization. At iteration $i=0$, set $\hat{\tau}^{(0)}(t)=1,\ t=1,\dots,T$, and $\hat{\boldsymbol{\Sigma}}^{(0)}_\text{n}=1/N\cdot\boldsymbol{I}_N$.
\item \textbf{Step 2}: Calculate $\hat{\Delta}^{(i)}$ from Eq.~(\ref{Delta_es}) using $\hat{\tau}^{(i)}(t)$ and $\hat{\boldsymbol{\Sigma}}^{(i)}_\text{n}$, then $\hat{\boldsymbol{\alpha}}^{(i)}$ from Eq.~(\ref{alpha_es}) using $\hat{\Delta}^{(i)}$, $\hat{\tau}^{(i)}(t)$ and $\hat{\boldsymbol{\Sigma}}^{(i)}_\text{n}$, and finally $\hat{\boldsymbol{v}}^{(i)}(t)$ from Eq.~(\ref{v}) using $\hat{\Delta}^{(i)}$ and $\hat{\boldsymbol{\alpha}}^{(i)}$.
\item \textbf{Step 3}: Use $\hat{\boldsymbol{v}}^{(i)}(t)$ and $\hat{\boldsymbol{\Sigma}}^{(i)}_\text{n}$ to update $\hat{\boldsymbol{\Sigma}}^{(i+1)}_\text{n}$ from Eqs.~(\ref{Sig_es2}) and (\ref{Sig_es_n}). Then, use $\hat{\boldsymbol{v}}^{(i)}(t)$ and the updated $\hat{\boldsymbol{\Sigma}}^{(i+1)}_\text{n}$ to find the updated $\hat{\tau}^{(i+1)}(t)$ from Eq.~(\ref{tau_es}). Set $i=i+1$.
\end{itemize}
Repeat Step 2 and Step 3 until a stop criterion (convergence or a maximum number of iteration) to obtain the final estimate of $\Delta$, which is denoted by $\hat{\Delta}_\text{IMLE}$.

The following remarks on our IMLE are in order:

\noindent\textbf{Remark 1}: The \emph{convergence} of the LL function in our algorithm is guaranteed by the fact that the value of the objective function at each step can either improve or maintain but cannot increase\cite{vorobyov1}. In fact, as the simulations in Section \ref{sec6} show, the convergence of the estimate of the unknown parameter $\Delta$ can also be observed with only two iteration, a result in accordance with those in \cite{PG01} and \cite{vorobyov1}. Here the convergence of $\Delta$ is defined as that $\vecnm{\hat{\Delta}^{(i+1)}-\hat{\Delta}^{(i)}}$ falls into a small range $\epsilon$, and further iterations do not lead to substantial improvement of performance in terms of the resulting mean square errors (MSEs).

\noindent\textbf{Remark 2}: Based on the observation in Remark 1, we can conclude that the \emph{computational cost} of our algorithm, which lies mainly in the solution of the highly nonlinear optimization problem in Step 2, is only a few times of that of the conventional MLE (CMLE). The latter corresponds to the case where the clutter is assumed to be uniform white Gaussian, such that Eq.~(\ref{Delta_es}) degenerate into:
\begin{equation}\label{conv_ml}
\hat{\Delta}_\text{CMLE}= \arg \min_{\Delta}\left\{\vecnm{\boldsymbol{\Pi}_{{\boldsymbol{B}}(\Delta)}^{\bot}{\boldsymbol{y}}}\right\}.
\end{equation}

\noindent\textbf{Remark 3}: One should also notice that, in the case where $T<N$, the sample covariance matrix is rank deficient. In this case, the Moore-Penrose pseudoinverse rather than the true inverse should be used for the calculation of $\boldsymbol{\Sigma}^{-1}$ in Eqs.~(\ref{tau_es}) and (\ref{Sig_es2}), as well as of $\boldsymbol{G}^{-1/2}$ in the expression of $\check{\boldsymbol{B}}(\Delta)$ and $\check{\boldsymbol{y}}$.

\section{Iterative Maximum A Posteriori Estimator}\label{sec3b}
The IMLE presented in Section \ref{sec3}, in which we treat the texture as deterministic and thereby ignore information regarding its statistical properties, has the advantage of easier and faster implementation. It is also a natural approach when the texture does not have a closed-form expression of distribution (e.g., in the case of Weibull clutter) or its distribution is unknown. In general cases, however, such approach is \emph{suboptimal}. In this section, we propose the IMAPE, which is also based on the idea of numerical concentration. Nevertheless, unlike the IMLE, the proposed IMAPE exploits information from the texture's prior distribution and leads to superior performance.

The maximum a posteriori estimator maximizes the joint LL function, denoted by $\Lambda_\text{J}$, which is equal to:
\begin{equation} \label{ll_joi}
\begin{aligned}
\Lambda_\text{J}&=\ln p_{\boldsymbol y, \boldsymbol{\tau}}\left(\boldsymbol y,  \boldsymbol{\tau}; \boldsymbol{\xi} \right)
=
\ln \left(p_{\boldsymbol y| \boldsymbol{\tau}}\left(\boldsymbol y | \boldsymbol{\tau}; \boldsymbol{\psi} \right)
p_{\boldsymbol \tau}(\boldsymbol \tau;a,b)\right)\\
&=\Lambda_\text{C}+\sum_{t=1}^T \ln p_{\tau(t)}(\tau(t);a,b)\\
&=\left\{
\begin{aligned}
&\Lambda_\text{C}-T\ln\Gamma(a)-Ta\ln{b}+(a-1)\sum_{t=1}^{T}\ln\tau(t)\\
&-\frac{\sum_{t=1}^{T}\tau(t)}{b}, \quad \text{K-distributed clutter},\\
&\Lambda_\text{C}-T\ln\Gamma(a)+Ta\ln{b}-(a+1)\sum_{t=1}^{T}\ln\tau(t)\\
&-b\sum_{t=1}^{T}\frac{1}{\tau(t)}, \quad \text{t-distributed clutter}.
\end{aligned}
\right.
\end{aligned}
\end{equation}

The expression of $\hat{\tau}(t)$, when all the remaining unknown parameters are fixed, can be found by solving $\partial\Lambda_\text{J}/\partial\tau(t)=0$, as:
\begin{equation}\label{tau_es2}
\hat{\tau}(t)=\left\{
\begin{aligned}
&\frac{1}{2}\bigg(\left(a-N-1\right)b+\Big(\left(a-N-1\right)^2b^2\\
&+4b
\left(\boldsymbol{y}(t)-\boldsymbol{v}(t)\right)^H\boldsymbol{\Sigma}^{-1}
\left(\boldsymbol{y}(t)-\boldsymbol{v}(t)\right)\Big)^\frac{1}{2}\bigg),\\
&\qquad\qquad\qquad\qquad\qquad    \text{K-distributed clutter},\\
&\frac{\left(\boldsymbol{y}(t)-\boldsymbol{v}(t)\right)^H\boldsymbol{\Sigma}^{-1}
\left(\boldsymbol{y}(t)-\boldsymbol{v}(t)\right)+b}{a+N+1},\\
&\qquad\qquad\qquad\qquad\qquad   \text{t-distributed clutter}.
\end{aligned}
\right.
\end{equation}

A comparison between the expressions of $\hat{\tau}(t)$ in Eq.~(\ref{tau_es}) and in Eq.~(\ref{tau_es2}) reveals that the latter takes into account the statistical properties of the texture. In these expressions, the parameters $a$ and $b$ play the roles of scale/translation factors to enhance the estimation of $\tau(t)$. This is more easily perceptible in the case of a t-distributed clutter, where the expressions for $\hat{\tau}(t)$ in Eq.~(\ref{tau_es}) and (\ref{tau_es2}) have a similar form. For example, the case of large $b$ and small $a$ corresponds to a more heavily-tailed distribution of the texture. This leads to an increased probability of the realization of $\tau(t)$ with large values. We note that the estimator in Eq.~(\ref{tau_es2}), in contrast to that in Eq.~(\ref{tau_es}), adjusts $\hat{\tau}(t)$ in a way that prevents the occurrence of small values and encourages that of larger ones.

Next, we consider the estimate of the texture parameters $a$ and $b$, denoted by $\hat{a}$ and $\hat{b}$. The latter can be obtained by solving $\partial\Lambda_\text{J}/\partial b=0$, as:
\begin{equation}\label{b_es}
\hat{b}=\left\{
\begin{aligned}
&\frac{\sum_{t=1}^{T}\tau(t)}{Ta}, \quad \text{K-distributed clutter},\\
&\frac{Ta}{\sum_{t=1}^{T}\frac{1}{\tau(t)}}, \quad \text{t-distributed clutter}.
\end{aligned}
\right.
\end{equation}On the other hand, calculating $\partial\Lambda_\text{J}/\partial a$ yields:
\begin{equation}\label{a_es}
\frac{\partial\Lambda_\text{J}}{\partial a}=\left\{
\begin{aligned}
&-T\Psi(a)-T\ln b+\sum_{t=1}^{T}\ln\tau(t), \quad \text{K-distributed clutter},\\
&-T\Psi(a)+T\ln b-\sum_{t=1}^{T}\ln\tau(t), \quad \text{t-distributed clutter},
\end{aligned}
\right.
\end{equation}in which $\Psi(\cdot)$ stands for the digamma function. From Eq.~(\ref{a_es}) it turns out that $\partial\Lambda_\text{J}/\partial a=0$ does not allow an analytical expression of the root, thus $\hat{a}$, unlike $\hat{b}$ in Eq.~(\ref{b_es}), can only be calculated numerically. Eqs.~(\ref{tau_es2})-(\ref{a_es}) reveal that the estimates of $\tau(t)$, $a$ and $b$ are mutually dependent, and further dependent on the parameter vector $\boldsymbol{\psi}$.

Now, let us approach the estimation of the target parameters and the speckle covariance matrix. The same expressions of $\hat{\boldsymbol{\Sigma}}$ and $\hat{\boldsymbol{\alpha}}$ in Eqs.~(\ref{Sig_es}) and (\ref{alpha_es}), that we obtained for the IMLE, are also valid in the case of the IMAPE, because $\partial\Lambda_\text{J}/\partial {\boldsymbol{\Sigma}}=\partial\Lambda_\text{C}/\partial {\boldsymbol{\Sigma}}$ and $\partial\Lambda_\text{J}/\partial {\boldsymbol{\alpha}}=\partial\Lambda_\text{C}/\partial {\boldsymbol{\alpha}}$. Substituting $\hat{\tau}(t)$ in Eq.~(\ref{tau_es2}) into Eq.~(\ref{Sig_es}), we arrive at the following iterative expression for $\hat{\boldsymbol{\Sigma}}$:
\begin{equation}\label{Sig_es4}
\hat{\boldsymbol{\Sigma}}^{(i+1)}=\left\{
\begin{aligned}
&\frac{2}{T}\sum_{t=1}^{T}\left(\boldsymbol{y}(t)-\boldsymbol{v}(t)\right)
\left(\boldsymbol{y}(t)-\boldsymbol{v}(t)\right)^H\\
&\left.\middle/\Bigg(\right.
\bigg(4b
\left(\boldsymbol{y}(t)-\boldsymbol{v}(t)\right)^H\left(\hat{\boldsymbol{\Sigma}}^{(i)}\right)^{-1}\\
&\cdot\left(\boldsymbol{y}(t)-\boldsymbol{v}(t)\right)
+\left(a-N-1\right)^2b^2
\bigg)^\frac{1}{2}\\
&+\left(a-N-1\right)b\Bigg),\quad \text{K-distributed clutter},\\
&\frac{a+N+1}{T}\sum_{t=1}^{T}
\Big(\left(\boldsymbol{y}(t)-\boldsymbol{v}(t)\right)\\
&\cdot\left(\boldsymbol{y}(t)-\boldsymbol{v}(t)\right)^H\Big)\\
&\left.\middle/\bigg(\right.\left(\boldsymbol{y}(t)-\boldsymbol{v}(t)\right)^H
\left(\hat{\boldsymbol{\Sigma}}^{(i)}\right)^{-1}\\
&\cdot\left(\boldsymbol{y}(t)-\boldsymbol{v}(t)\right)+b\bigg),\quad\text{t-distributed clutter},
\end{aligned}
\right.
\end{equation}which, similar to the expression of $\hat{\boldsymbol{\Sigma}}^{(i+1)}$ in Eq.~(\ref{Sig_es2}) for the IMLE, needs to be substituted into Eq.~(\ref{Sig_es_n}) to obtain the normalized $\hat{\boldsymbol{\Sigma}}^{(i+1)}$ denoted as $\hat{\boldsymbol{\Sigma}}^{(i+1)}_\text{n}$.

Finally, we address the estimation of $\Delta$. Adopting the numerical concentration approach similar to that in Section \ref{sec3}, we assume here that $\boldsymbol{\Sigma}$ and $\boldsymbol{\tau}$ to be known from the previous iteration of the algorithm. Furthermore, as the estimates of $a$ and $b$ are only dependent on $\boldsymbol{\tau}$, these are also fixed for each iteration. Thus, we may drop in the expression of the joint LL function $\Lambda_\text{J}$ in Eq.~(\ref{ll_joi}) those terms that contain only these parameters, transforming it into the same expression as in Eq.~(\ref{ll_con2}). This means that $\Delta$ can be obtained, also for the IMAPE, from Eq.~(\ref{Delta_es}).

The iterative estimation procedure of the proposed IMAPE also contains three steps and is summarized as follows:
\begin{itemize}
\item \textbf{Step 1}: Initialization. At iteration $i=0$, set $\hat{\tau}^{(0)}(t)=1,\ t=1,\dots,T$, and $\hat{\boldsymbol{\Sigma}}^{(0)}_\text{n}=1/N\cdot\boldsymbol{I}_N$.
\item \textbf{Step 2}: Calculate $\hat{\Delta}^{(i)}$ from Eq.~(\ref{Delta_es}) using $\hat{\tau}^{(i)}(t)$ and $\hat{\boldsymbol{\Sigma}}^{(i)}_\text{n}$, then $\hat{\boldsymbol{\alpha}}^{(i)}$ from Eq.~(\ref{alpha_es}) using $\hat{\Delta}^{(i)}$, $\hat{\tau}^{(i)}(t)$ and $\hat{\boldsymbol{\Sigma}}^{(i)}_\text{n}$. Next, calculate $\hat{\boldsymbol{v}}^{(i)}(t)$ from Eq.~(\ref{v}) using $\hat{\Delta}^{(i)}$ and $\hat{\boldsymbol{\alpha}}^{(i)}$. Finally, substitute Eq.~(\ref{b_es}) into Eq.~(\ref{a_es}), and find numerically $\hat{a}^{(i)}$ from Eq.~(\ref{a_es}) using $\hat{\tau}^{(i)}(t)$, then find $\hat{b}^{(i)}$ from Eq.~(\ref{b_es}) using $\hat{\tau}^{(i)}(t)$ and $\hat{a}^{(i)}$.
\item \textbf{Step 3}: Use $\hat{\boldsymbol{v}}^{(i)}(t)$, $\hat{\boldsymbol{\Sigma}}^{(i)}_\text{n}$, $\hat{a}^{(i)}$ and $\hat{b}^{(i)}$ to update $\hat{\boldsymbol{\Sigma}}^{(i+1)}_\text{n}$ from Eqs.~(\ref{Sig_es4}) and (\ref{Sig_es_n}). Then, use $\hat{\boldsymbol{v}}^{(i)}(t)$, $\hat{a}^{(i)}$, $\hat{b}^{(i)}$ and the updated $\hat{\boldsymbol{\Sigma}}^{(i+1)}_\text{n}$ to find the updated $\hat{\tau}^{(i+1)}(t)$ from Eq.~(\ref{tau_es2}). Set $i=i+1$.
\end{itemize}
Repeat Step 2 and Step 3 until a stop criterion (convergence or a maximum number of iteration) to obtain the final $\hat{\Delta}$, denoted by $\hat{\Delta}_\text{IMAPE}$.

Note that Remarks 1-3 of Section \ref{sec3} also directly apply to the proposed IMAPE.

\section{Cram\'{e}r-Rao-like bounds}\label{sec4}
The CRLBs provide an essential tool for evaluating the performance of any unbiased estimator. Furthermore, closed-form expressions of the CRLBs are required in the computation of the expression for the ARL in Smith's sense. In this section, we derive the expressions of various CRLBs w.r.t. $\Delta$, including the standard CRB, the EMCB, the MCRB and the HCRB, and provides a comparison between them.

\subsection{Standard Cram\'{e}r-Rao Bound}\label{subsec_scrb}
In \cite{xzh3}, we have derived the expression for the standard CRB w.r.t. $\Delta$, denoted by $\text{CRB}(\Delta)$, under a K-distributed clutter. This result also holds true for the t-distributed clutter case, except for the factor $\kappa$ (that will be detailed later), which takes another expression under a t-distributed clutter.

$\text{CRB}(\Delta)$ considers the parameter vector $\boldsymbol{\xi}$, and is obtained as the upper-leftmost element of the inverse of the Fisher Information Matrix (FIM), denoted by $\boldsymbol{F}$. The FIM is calculated from the marginal likelihood $p_{\boldsymbol y}\left(\boldsymbol y; \boldsymbol{\xi} \right)$ in Eq.~(\ref{n1b}). The elements of $\boldsymbol{F}$ are given by:
\begin{equation}\label{fim_s}
\left[\boldsymbol{F}\right]_{i,j}
=\text{E}_{\boldsymbol{y}}\left\{\frac{\partial\ln\left(p_{\boldsymbol y}\left(\boldsymbol y; \boldsymbol{\xi} \right)\right)}{\partial\left[\boldsymbol{\xi}\right]_{i}}\frac{\partial\ln\left(p_{\boldsymbol y}\left(\boldsymbol y; \boldsymbol{\xi} \right)\right)}{\partial\left[\boldsymbol{\xi}\right]_{j}}\right\},
\end{equation}in which $[\cdot]_{i,j}$ denotes the $(i,j)$th entry of a matrix, and $[\cdot]_i$ denotes the $i$th element of a vector. Derivations show that $\boldsymbol{F}$ takes the following block-diagonal structured form:
\begin{equation}\label{7aa}
\boldsymbol{F}=\left[
\begin{array}{cc}
\boldsymbol{\Phi} & \boldsymbol{0}_{5\times (N^2+2)} \\
\boldsymbol{0}_{(N^2+2) \times 5 } & \boldsymbol{\Xi}
\end{array}\right],
\end{equation}in which $\boldsymbol{\Phi}$ denotes the $5 \times 5$ FIM block w.r.t. the target parameters (those in $\boldsymbol{\mu}$), whose entries, denoted by $\phi_{ij}$, are given by:
\begin{equation}\label{f_S}
\phi_{ij}=\frac{2\kappa}{N}\sum_{t=1}^{T}\overline{\text{tr}\left\{\boldsymbol{v}_{i}(t)
\boldsymbol{v}^H_{j}(t)\boldsymbol{\Sigma}^{-1}\right\}},\ i,j=1,\dots,5,
\end{equation}where $\boldsymbol{v}_{i}(t)=\partial\boldsymbol{v}(t)/\partial\left[\boldsymbol{\mu}\right]_i$. The matrix $\boldsymbol{\Xi}$ in Eq.~(\ref{7aa}) represents the FIM block w.r.t. the clutter parameters ($a$, $b$ and $\left[\boldsymbol{\zeta}\right]_{i}$). As $\boldsymbol{\Phi}$ and $\boldsymbol{\Xi}$ are decoupled, we have:
\begin{equation}\label{SCRB}
\text{CRB}\left(\Delta\right)=\left[\boldsymbol{\Phi}^{-1}\right]_{1,1}.
\end{equation}

The expression of the positive real factor $\kappa$ in Eq.~(\ref{f_S}) depends on the distribution of the texture and is given by:
\begin{equation}\label{kappa}
\kappa=\left\{
\begin{aligned}
&\frac{\int_0^{+\infty}x^{N+a-1}\frac{K_{a-N-1}^2(x)}
{K_{a-N}(x)}\text{d}x}{2^{N+a-2}b\Gamma(N)\Gamma(a)}, \quad \text{K-distributed clutter}, \\
&\frac{Na(a+N)}{b(a+N+1)},\quad \text{t-distributed clutter};
\end{aligned}
\right.
\end{equation}in which $K_n(x)$ is the modified Bessel functions of the second kind of order $n$. For a t-distributed clutter, Eq.~(\ref{kappa}) is a generalization of the result in \cite{mim15} to the two texture parameter cases. For a K-distributed clutter, we have found a more compact expression of $\kappa$ than \cite{mim15}, which yet still can only be evaluated numerically.

\subsection{Extended Miller-Chang Bound}
The EMCB was first proposed in \cite{gini2} as an extension to the conventional Miller-Chang Bound (MCB) \cite{miller1}. Its general motivation is to first treat the random nuisance parameters ($\boldsymbol \tau$ in our case) as deterministic and derive the CRB calculated from the conditional likelihood $p_{\boldsymbol y| \boldsymbol{\tau}}\left(\boldsymbol y | \boldsymbol{\tau}; \boldsymbol{\psi}\right)$ in Eq.~(\ref{n1}). Then in the next step, the assumption of constant $\boldsymbol \tau$ is relaxed and the CRB is averaged over different realizations of $\boldsymbol \tau$ drawn from the corresponding random distribution. This approach has in common with the proposed IMLE in Section \ref{sec3}, that the latter also treats $\boldsymbol \tau$ to be deterministic. The performance of this algorithm, in terms of the averaged MSE resulting from many independent Monte-Carlo trials, can be evaluated by averaging the CRBs calculated for each of the trials. It is clear that such an averaged CRB, when the trial number becomes large, approaches the EMCB.

The parameter vector $[\boldsymbol{\psi}^T, \boldsymbol{\tau}^T]^T$ is considered in the calculation of the EMCB. The entries of the corresponding FIM, denoted by $\boldsymbol{F}_{\text{E}}$, are calculated by:
\begin{equation}\label{fim_e}
\left[\boldsymbol{F}_{\text{E}}\right]_{i,j}=\text{E}_{\boldsymbol{y}|\boldsymbol{\tau}}
\left\{\frac{\partial\ln\left(p_{\boldsymbol y| \boldsymbol{\tau}}\left(\boldsymbol y|  \boldsymbol{\tau}; \boldsymbol{\psi} \right)\right)}{\partial\left[\boldsymbol{\psi}\right]_{i}}\frac{\partial\ln\left(p_{\boldsymbol y| \boldsymbol{\tau}}\left(\boldsymbol y,|  \boldsymbol{\tau}; \boldsymbol{\psi} \right)\right)}{\partial\left[\boldsymbol{\psi}\right]_{j}}\right\}.
\end{equation}whose calculation resembles that of the FIM under Gaussian clutter (with the difference that the data are weighted by $1/\tau(t)$ varying at each snapshot) and is omitted here for brevity. Similar to $\boldsymbol{F}$, $\boldsymbol{F}_{\text{E}}$ exhibits a block-diagonal structure, where the blocks for the target and clutter parameters are decoupled from each other. We denote the parameter block of interest by $\boldsymbol{\Phi}_\text{E}$, and its entries by $\phi^{\text{E}}_{ij},\ i,j=1,\dots,5$. The following expressions are obtained:
\begin{equation}\label{f_E}
\phi^{\text{E}}_{ij}=
2\sum_{t=1}^{T}\frac{1}{\tau(t)}\overline{\text{tr}\left\{\boldsymbol{v}_{i}(t)
\boldsymbol{v}^H_{j}(t)\boldsymbol{\Sigma}^{-1}\right\}}.
\end{equation}Consequently, the EMCB w.r.t. $\Delta$ , denoted by $\text{EMCB}\left(\Delta\right)$, is given by:
\begin{equation}\label{EMCB}
\text{EMCB}\left(\Delta\right)=\text{E}_{\boldsymbol \tau}\left\{\left[\boldsymbol{\Phi}_\text{E}^{-1}\right]_{1,1}\right\},
\end{equation}for which no closed-form expression exists.

\subsection{Modified and Hybrid Cram\'{e}r-Rao Bound}
The MCRB \cite{gini3}, like the EMCB, also considers the unknown parameter vector as $[\boldsymbol{\psi}^T, \boldsymbol{\tau}^T]^T$. Its corresponding FIM, denoted by $\boldsymbol{F}_{\text{M}}$, is likewise calculated from the conditional likelihood in Eq.~(\ref{n1}). The MCRB differs from the EMCB only in that it averages over the random parameters before the FIM inversion, namely:
\begin{equation}\label{fim_m}
\begin{aligned}
&\left[\boldsymbol{F}_{\text{M}}\right]_{i,j}=\text{E}_{\boldsymbol{y},\boldsymbol{\tau}}
\left\{\frac{\partial\ln\left(p_{\boldsymbol y| \boldsymbol{\tau}}\left(\boldsymbol y|  \boldsymbol{\tau}; \boldsymbol{\psi} \right)\right)}{\partial\left[\boldsymbol{\psi}\right]_{i}}\frac{\partial\ln\left(p_{\boldsymbol y| \boldsymbol{\tau}}\left(\boldsymbol y,|  \boldsymbol{\tau}; \boldsymbol{\psi} \right)\right)}{\partial\left[\boldsymbol{\psi}\right]_{j}}\right\}\\
&=\text{E}_{\boldsymbol{\tau}}\left\{\text{E}_{\boldsymbol{y}|\boldsymbol{\tau}}
\left\{\frac{\partial\ln\left(p_{\boldsymbol y| \boldsymbol{\tau}}\left(\boldsymbol y|  \boldsymbol{\tau}; \boldsymbol{\psi} \right)\right)}{\partial\left[\boldsymbol{\psi}\right]_{i}}\frac{\partial\ln\left(p_{\boldsymbol y| \boldsymbol{\tau}}\left(\boldsymbol y,|  \boldsymbol{\tau}; \boldsymbol{\psi} \right)\right)}{\partial\left[\boldsymbol{\psi}\right]_{j}}\right\}\right\}\\
&=\text{E}_{\boldsymbol{\tau}}\left\{\left[\boldsymbol{F}_{\text{E}}\right]_{i,j}\right\}.
\end{aligned}
\end{equation}Similar to $\boldsymbol{F}$ and $\boldsymbol{F}_{\text{E}}$, $\boldsymbol{F}_{\text{M}}$ also has a block-diagonal structure, whose parameter block of interest, denoted by $\boldsymbol{\Phi}_\text{M}$, contains the following entries $\phi^{\text{M}}_{ij},\ i,j=1,\dots,5$:
\begin{equation}\label{f_M}
\phi^{\text{M}}_{ij}
=\text{E}_{\boldsymbol{\tau}}\left\{\left[\boldsymbol{\Phi}_\text{E}^{-1}\right]_{i,j}\right\}
=2\nu\sum_{t=1}^{T}\overline{\text{tr}\left\{\boldsymbol{v}_{i}(t)
\boldsymbol{v}^H_{j}(t)\boldsymbol{\Sigma}^{-1}\right\}},
\end{equation}in which
\begin{equation}\label{nu}
\nu=\text{E}\left\{\frac{1}{\tau(t)}\right\}
=\left\{
\begin{aligned}
&\frac{2}{b(a-1)},\quad \text{K-distributed clutter, for $a>1$}, \\
&\frac{2a}{b},\quad \text{t-distributed clutter},
\end{aligned}
\right.
\end{equation}and the MCRB w.r.t. $\Delta$, denoted by $\text{MCRB}\left(\Delta\right)$, is equal to:
\begin{equation}\label{MCRB}
\text{MCRB}\left(\Delta\right)=\left[\boldsymbol{\Phi}_\text{M}^{-1}\right]_{1,1}.
\end{equation}

The HCRB as defined in \cite{reuven1}, on the other hand, considers the unknown parameter vector as $[\boldsymbol{\xi}^T, \boldsymbol{\tau}^T]^T$. Furthermore, it uses the joint likelihood in Eq.~(\ref{n1a}), instead of the conditional likelihood in Eq.~(\ref{n1}), similar as in the derivation of the EMCB and MCRB, to obtain its FIM, which is denoted by $\boldsymbol{F}_{\text{H}}$. The entries of $\boldsymbol{F}_{\text{H}}$ are calculated by:
\begin{equation}\label{fim_h}
\phi^{\text{H}}_{ij}=\text{E}_{\boldsymbol{y},\boldsymbol{\tau}}
\left\{\frac{\partial\ln\left(p_{\boldsymbol y, \boldsymbol{\tau}}\left(\boldsymbol y,  \boldsymbol{\tau}; \boldsymbol{\xi} \right)\right)}{\partial\left[\boldsymbol{\xi}\right]_{i}}\frac{\partial\ln\left(p_{\boldsymbol y, \boldsymbol{\tau}}\left(\boldsymbol y,  \boldsymbol{\tau}; \boldsymbol{\xi} \right)\right)}{\partial\left[\boldsymbol{\xi}\right]_{j}}\right\}.
\end{equation}Our derivations show that $\boldsymbol{F}_{\text{H}}$ also has a block structure, and its parameter block of interest is equal to that of the MCRB, $\boldsymbol{\Phi}_\text{M}$. Consequently, we have:
\begin{equation}\label{HCRB}
\text{HCRB}\left(\Delta\right)=\text{MCRB}\left(\Delta\right),
\end{equation}in which $\text{HCRB}\left(\Delta\right)$ represents the HCRB w.r.t. $\Delta$.

\subsection{Relationships between the CRLBs}\label{relation}
It is theoretically proved in \cite{reuven1} that the standard CRB is always larger than the HCRB. As we also have $\text{HCRB}\left(\Delta\right)=\text{MCRB}\left(\Delta\right)$, it follows that:
\begin{equation}\label{crb_re1}
\text{CRB}\left(\Delta\right)\geq\text{HCRB}\left(\Delta\right)=\text{MCRB}\left(\Delta\right).
\end{equation}This relationship, however, becomes apparent when the clutter follows a t-distribution, where $\text{CRB}\left(\Delta\right)$ has a closed-form expression. By comparison of Eqs.~(\ref{f_S})-(\ref{kappa}) with Eqs.~(\ref{SCRB}), (\ref{f_M}) and (\ref{HCRB}), we have:
\begin{equation}
\frac{\text{CRB}\left(\Delta\right)}{\text{MCRB}\left(\Delta\right)}=
\frac{\text{CRB}\left(\Delta\right)}{\text{HCRB}\left(\Delta\right)}=
\frac{a+N+1}{a+N}>1.
\end{equation}Moreover, since $(a+N+1)/(a+N)\rightarrow1$ when $N\rightarrow\infty$, it follows that $\text{CRB}(\Delta)\rightarrow\text{MCRB}(\Delta)=\text{HCRB}(\Delta)$ when the number of receiver antennas becomes large.

The relationship between $\text{EMCB}(\Delta)$ and $\text{MCRB}\left(\Delta\right)$ (or $\text{HCRB}\left(\Delta\right)$) can be revealed by noticing, from Eq.~(\ref{EMCB}), that:
\begin{equation}
\text{EMCB}(\Delta)
=\text{E}_{\boldsymbol \tau}\left\{\left[\boldsymbol{\Phi}_\text{E}^{-1}\right]_{1,1}\right\}
=\left[\text{E}_{\boldsymbol \tau}\left\{\boldsymbol{\Phi}_\text{E}^{-1}\right\}\right]_{1,1},
\end{equation}and, according to Eqs.~(\ref{fim_m}) and (\ref{MCRB}), that:
\begin{equation}
\text{MCRB}(\Delta)=\left[\boldsymbol{\Phi}_\text{M}^{-1}\right]_{1,1}
=\left[\left(\text{E}_{\boldsymbol \tau}\left\{\boldsymbol{\Phi}_\text{E}\right\}\right)^{-1}\right]_{1,1}.
\end{equation}Since $\boldsymbol{\Phi}_\text{E}^{-1}$ is a convex function of the entries of $\boldsymbol{\Phi}_\text{E}$ \cite{groves1}, by Jensen's inequality, we have:
\begin{equation}
\text{E}_{\boldsymbol \tau}\left\{\boldsymbol{\Phi}_\text{E}^{-1}\right\}-\left(\text{E}_{\boldsymbol \tau}\left\{\boldsymbol{\Phi}_\text{E}\right\}\right)^{-1}\succeq\boldsymbol{0},
\end{equation}Hence $\left[\text{E}_{\boldsymbol \tau}\left\{\boldsymbol{\Phi}_\text{E}^{-1}\right\}\right]_{1,1}\geq\left[\left(\text{E}_{\boldsymbol \tau}\left\{\boldsymbol{\Phi}_\text{E}\right\}\right)^{-1}\right]_{1,1}$, viz.,
\begin{equation}
\text{EMCB}(\Delta)\geq\text{MCRB}(\Delta)=\text{HCRB}(\Delta).
\end{equation}Furthermore, since $\boldsymbol{\Phi}_\text{E}\rightarrow\boldsymbol{\Phi}_\text{M}$ when $T\rightarrow\infty$, we have that $\text{EMCB}(\Delta)\rightarrow\text{MCRB}(\Delta)=\text{HCRB}(\Delta)$ as the number of snapshots becomes large.

The relationship between $\text{CRB}(\Delta)$ and $\text{EMCB}(\Delta)$, on the other hand, is indefinite and dependent on $T$ and $N$, as will be illustrated by numerical simulations.

\subsection{CRLBs and the Texture Parameters}\label{crlb_para}
At the end of this section, we investigate the impact of the clutter's texture parameters, $a$ and $b$, on the CRLBs. To achieve this, we first define the signal-to-clutter ratio (SCR) as \cite{mim15}:
\begin{equation}\label{tcr}
\text{SCR}=\frac{\sum_{t=1}^{T}\vecnm{\boldsymbol{s}(t)}^2}
{T\text{E}\{\tau(t)\}\sigma^2\text{tr}\left\{\check{\boldsymbol{\Sigma}}\right\}},
\end{equation}in which $\text{E}\{\tau(t)\}$ is equal to $ab$ for a K-distributed clutter and $b/(a-1)$ for a t-distributed clutter (for $a>1$) \cite{pp95}. It then turns out that for a fixed SCR, we have:
\begin{equation}\label{a_prop}
\frac{1}{\sigma^2}\propto
\left\{
\begin{aligned}
&a, \quad \text{K-distributed clutter}, \\
&\frac{1}{a-1},\quad  \text{t-distributed clutter, for $a>1$},
\end{aligned}
\right.
\end{equation}and
\begin{equation}\label{b_prop}
\frac{1}{\sigma^2}\propto b, \quad \text{K-distributed and t-distributed clutters},
\end{equation}in which $\propto$ denotes direct proportionality. Furthermore, from Eq.~(\ref{kappa}), we have:
\begin{equation}\label{kappa_a_prop}
\kappa\propto
\left\{
\begin{aligned}
&\frac{\int_0^{+\infty}x^{N+a-1}\frac{K_{a-N-1}^2(x)}
{K_{a-N}(x)}\text{d}x}{2^{N+a-2}\Gamma(a)}, \quad \text{K-distributed clutter}, \\
&\frac{a(a+N)}{(a+N+1)},\quad  \text{t-distributed clutter, for $a>1$},
\end{aligned}
\right.
\end{equation}and
\begin{equation}\label{kappa_b_prop}
\kappa\propto \frac{1}{b}, \quad \text{K-distributed and t-distributed clutters}.
\end{equation}

\subsubsection{CRLBs vs. $a$}
We begin with the standard CRB. The expression in Eq.~(\ref{f_S}) can be converted to:
\begin{equation}\label{f_S2}
\phi_{ij}=\frac{2\kappa}{N\sigma^2}\sum_{t=1}^{T}\overline{\text{tr}\left\{\boldsymbol{v}_{i}(t)
\boldsymbol{v}^H_{j}(t)\check{\boldsymbol{\Sigma}}^{-1}\right\}},\ i,j=1,\dots,5,
\end{equation}namely, $\phi_{ij}\propto\kappa/\sigma^2$, to which we apply Eqs.~(\ref{kappa_a_prop}) and (\ref{a_prop}) and have straightforwardly:
\begin{equation}
\phi_{ij}\propto
\left\{
\begin{aligned}
&\frac{a\int_0^{+\infty}x^{N+a-1}\frac{K_{a-N-1}^2(x)}
{K_{a-N}(x)}\text{d}x}{2^{N+a-2}\Gamma(a)}, \quad \text{K-distributed clutter}, \\
&\frac{a(a+N)}{(a+N+1)(a-1)},\quad  \text{t-distributed clutter, for $a>1$}.
\end{aligned}
\right.
\end{equation}For both clutter distributions $\phi_{ij}$ decreases as $a$ increases\footnote{This relationship is obvious for t-distributed clutter, for K-distributed clutter, however, for which $\phi_{ij}$ does not enjoy a closed-form expression, can only be determined numerically.}; as a result, $\text{CRB}\left(\Delta\right)$ increases with $a$, i.e., the standard CRB is \emph{positively correlated} with the shape parameter $a$.

Similarly, we deduce from Eqs.~(\ref{nu}) and (\ref{a_prop}) that:
\begin{equation}
\phi^{\text{M}}_{ij}=\phi^{\text{H}}_{ij}\propto\frac{a}{a-1},\quad \text{K-distributed and t-distributed clutters},
\end{equation}also indicating a positive correlation between the MCRB/HCRB and $a$. Furthermore, we notice, as opposed to the standard CRB, which has different proportionalities to $a$ for K-distributed and t-distributed clutters respectively, the MCRB/HCRB have the same proportionality for both clutter distributions.

Finally, for the EMCB, we have from Eq.~(\ref{f_E}) that:
\begin{equation}\label{f_E}
\phi^{\text{E}}_{ij}=
\frac{2}{\sigma^2}\sum_{t=1}^{T}\frac{1}{\tau(t)}\overline{\text{tr}\left\{\boldsymbol{v}_{i}(t)
\boldsymbol{v}^H_{j}(t)\check{\boldsymbol{\Sigma}}^{-1}\right\}}
\propto\frac{1}{\sigma^2}\sum_{t=1}^{T}\frac{1}{\tau(t)}.
\end{equation}Consequently, it follows from Eq.~(\ref{EMCB}) that:
\begin{equation}\label{EMCB_prop}
\text{EMCB}\left(\Delta\right)=\sigma^2\text{E}_{\boldsymbol \tau}\left\{\frac{1}{\sum_{t=1}^{T}\frac{1}{\tau(t)}}\right\}.
\end{equation}For a t-distributed clutter, $\tau(t)\sim\text{Inv-Gamma}(a, b)$, $1/\tau(t)\sim\text{Gamma}(a, 1/b)$. Thus, as $\tau(t),\ t=1,\dots,T$ are i.i.d. variables, from the property of the gamma distribution arises that $1/\sum_{t=1}^{T}\left(1/\tau(t)\right)\sim\text{Gamma}(Ta, 1/b)$, and consequently,
\begin{equation}\label{E_tau_prop}
\text{E}_{\boldsymbol \tau}\left\{\frac{1}{\sum_{t=1}^{T}\frac{1}{\tau(t)}}\right\}=\frac{Ta}{b}, \quad \text{t-distributed clutter},
\end{equation}which, combined with Eqs.~(\ref{a_prop}) and (\ref{EMCB_prop}), results in $\text{EMCB}\left(\Delta\right)\propto a(a-1)$, indicating a positive correlation also between the EMCB and $a$. For a K-distributed clutter an analogous deduction seems, however, impossible or at least complicated, due to the presence of the sum of inverse gamma variables. The relationship between the EMCB and $a$ for this case can be numerically ascertained.

\subsubsection{CRLBs vs. $b$}
Associating Eq.~(\ref{kappa_b_prop}) with Eq.~(\ref{f_S2}), yields:
\begin{equation}
\phi_{ij}\propto\frac{1}{b\sigma^2}, \quad \text{K-distributed and t-distributed clutters}.
\end{equation}As from Eq.~(\ref{b_prop}) for both clutter distributions $b\propto1/\sigma^2$, $\phi_{ij}$ is thus independent of $b$, which means under a fixed SCR, changing $b$ does not give rise to any variation in the value of $\text{CRB}(\Delta)$. The same also holds true for the MCRB/HCRB for both clutter distributions, and can be established in a similar vein by considering $\nu$ instead of $\kappa$.

The independence of the EMCB of $b$ under t-distributed clutter is straightforwardly confirmable by combining Eqs.~(\ref{b_prop}), (\ref{EMCB_prop}) and (\ref{E_tau_prop}). However, under K-distributed clutter, the relationship between the EMCB and $b$ can only be determined numerically.

In summary, the performance of the estimation, in terms of the lowest achievable CRLBs, is only related to the shape parameter $a$ of the clutter, and decreases as $a$ becomes larger, and is independent of the scale paramter $b$. This will also be verified in Section \ref{sec6} by numerical simulations.

\section{Derivation of the ARL}\label{sec5}
In this section, we address the question of the target resolvability. In order to obtain an analytical expression for the ARL in Smith's sense, a closed-form (non-matrix) expression for $\text{CRB}(\Delta)$ is required. Our above derived $\text{CRB}(\Delta)$ in Eq.~(\ref{SCRB}), however, cannot be analytically inverted, due to the nonlinearity of our model in Eq.~(\ref{1b}) w.r.t. $\Delta$. To cope with this difficulty, we first linearize the model \cite{zhxtc8,zhxtc6,mim18,SM05a,AW08}, and rederive the FIM expression based on it which is feasible for analytical inversion. The ARL obtained from the linearized model approximates the exact ARL obtained from the original model.

\subsection{Model Linearization}\label{subsection1}
To linearize the model, we resort to the second order Taylor expansion around $\Delta=0$ in Eq.~(\ref{1b}). This step of approximation is justified by considering the fact that, in asymptotic cases, e.g., those of large SCR or sample size, in which the CRB is a tight bound, the ARL is always very small, i.e., the value of $\Delta$ corresponding to the ARL approaches zero ($\Delta\ll1$) \cite{SM04,LN07,AW08,ShaMil05,zhxtc10}\footnote{This is also supported by the fact that the ML estimator, and generally all high resolution estimators, have asymptotically an infinite resolution capability, leading to the ARL infinitely approaching to $0$ \cite{SN89,zhxtc9}\label{foot1}.}. The second order Taylor expansions of $\boldsymbol{a}_{\mathcal T}\left(\omega_{2}\right)$ and $\boldsymbol{a}_{\mathcal R}\left(\omega_{2}\right)$ are respectively given by:
\begin{subequations}
{\setlength\arraycolsep{0.1em}
\begin{eqnarray}
\boldsymbol{a}_{\mathcal T}\left(\omega_{2}\right)\approx\boldsymbol{a}_{\mathcal T}\left(\omega_{1}\right)+j\Delta\boldsymbol{\dot a}_{\mathcal T}\left(\omega_{1}\right)-\frac{\Delta^{2}}{2}\boldsymbol{\ddot a}_{\mathcal T}\left(\omega_{1}\right),\\
\boldsymbol{a}_{\mathcal R}\left(\omega_{2}\right)\approx\boldsymbol{a}_{\mathcal R}\left(\omega_{1}\right)+j\Delta\boldsymbol{\dot a}_{\mathcal R}\left(\omega_{1}\right)-\frac{\Delta^{2}}{2}\boldsymbol{\ddot a}_{\mathcal R}\left(\omega_{1}\right),
\end{eqnarray}
}\end{subequations}where $\boldsymbol{\dot a}_{\mathcal T}\left(\cdot\right)=\boldsymbol{a}_{\mathcal T}\left(\cdot\right)\odot\boldsymbol{d}_{\mathcal T}$,
$\boldsymbol{\dot a}_{\mathcal R}\left(\cdot\right)=\boldsymbol{a}_{\mathcal R}\left(\cdot\right)\odot\boldsymbol{d}_{\mathcal R}$, $\boldsymbol{\ddot a}_{\mathcal T}\left(\cdot\right)\triangleq\boldsymbol{a}_{\mathcal T}\left(\cdot\right)\odot\boldsymbol{d}_{\mathcal T}\odot\boldsymbol{d}_{\mathcal T}$, $\boldsymbol{\ddot a}_{\mathcal R}\left(\cdot\right)\triangleq\boldsymbol{a}_{\mathcal R}\left(\cdot\right)\odot\boldsymbol{d}_{\mathcal R}\odot\boldsymbol{d}_{\mathcal R}$, in which $\odot$ denotes the Hadamard product, and $\boldsymbol{d}_{\mathcal T}=[0,d_{\mathcal T},\dots, (M-1)d_{\mathcal T}]^T$, $\boldsymbol{d}_{\mathcal R}=[0, d_{\mathcal R},\dots, (N-1)d_{\mathcal R}]^T$. One can then approximate Eq.~(\ref{1b}) as (omitting all terms containing $\Delta^n,\ n>2$):
\begin{equation}\label{3}
\boldsymbol{y}(t)\approx\boldsymbol{C}(t)\boldsymbol{\eta}+\boldsymbol{n}(t),\quad t=1,\dots,T;
\end{equation}
where $\boldsymbol{\eta}=\left[\alpha_{1}+\alpha_{2} \quad j\alpha_{2}\Delta \quad -\alpha_{2}\Delta^{2}\right]^{T}$, $\boldsymbol{C}(t)=\left[\boldsymbol{\rho}_{1}(t) \quad \boldsymbol{\rho}_{2}(t) \quad \boldsymbol{\rho}_{3}(t)\right]$, in which $\boldsymbol{\rho}_{i}(t)=\boldsymbol{R}_i\boldsymbol{s}(t),\ i,j=1,2,3$, where
\begin{subequations}
{\setlength\arraycolsep{0.1em}
\begin{eqnarray}
\boldsymbol{R}_{1}&=&\boldsymbol{a}_{\mathcal R}\left(\omega_{1}\right)\boldsymbol{a}_{\mathcal T}^{T}\left(\omega_{1}\right),\label{rho1}\\
\boldsymbol{R}_{2}&=&\boldsymbol{\dot a}_{\mathcal R}\left(\omega_{1}\right)\boldsymbol{a}_{\mathcal T}^{T}\left(\omega_{1}\right)+ \boldsymbol{a}_{\mathcal R}\left(\omega_{1}\right)\boldsymbol{\dot a}_{\mathcal T}^{T}\left(\omega_{1}\right),\label{rho2}\\
\boldsymbol{R}_{3}&=&\boldsymbol{\dot a}_{\mathcal R}\left(\omega_{1}\right)\boldsymbol{\dot a}_{\mathcal T}^{T}\left(\omega_{1}\right)
+\frac{1}{2} \boldsymbol{\ddot a}_{\mathcal R}\left(\omega_{1}\right)\boldsymbol{a}_{\mathcal T}^{T}\left(\omega_{1}\right)\nonumber\\
&&+\frac{1}{2} \boldsymbol{a}_{\mathcal R}\left(\omega_{1}\right)\boldsymbol{\ddot a}_{\mathcal T}^{T}\left(\omega_{1}\right).\label{rho3}
\end{eqnarray}
}\end{subequations}

\subsection{Analytical Expression of $\text{CRB}(\Delta)$}\label{ana_crb}
We obtain the analytical expression for $\text{CRB}\left(\Delta\right)$ by rederiving the FIM expression based on the model Eq.~(\ref{3}) and then invert its $5\times 5$ parameter block of interest. The procedure of the derivation, which can be found in Appendix \ref{appendix}, leads to the following result:
\begin{equation}\label{14}
\text{CRB}\left(\Delta\right)=\frac{1}{\phi_{11}'+Q},
\end{equation}for $Q=(\phi_{44}'\phi_{12}'^2+\phi_{44}'\phi_{13}'^2+\phi_{22}'\phi_{14}'^2+\phi_{22}'\phi_{15}'^2
-2\phi_{24}'\phi_{12}'\phi_{14}'-2\phi_{25}'\phi_{12}'\phi_{15}'+2\phi_{25}'\phi_{13}'\phi_{15}'-2\phi_{24}'\phi_{13}'
\phi_{15}')/(\phi_{24}'^2+\phi_{25}'^2-\phi_{22}'\phi_{44}')$, and in which $\phi_{ij}',\ i,j=1,\dots5$ are the entries of the parameter block of interest of the FIM based on the linearized model, defined in Eqs.~(\ref{f11})-(\ref{f22}).

By the same vein of the derivation procedure for Eq.~(\ref{14}), we can also obtain an analytical expression for $\text{MCRB}\left(\Delta\right)$ and $\text{HCRB}\left(\Delta\right)$. The resulting MCRB and HCRB retain the same expression as Eq.~(\ref{14}), yet with $\phi_{ij}'$ calculated by replacing $\kappa / N$ with $\nu$ in Eqs.~(\ref{f11})-(\ref{f22}). The analytical expression for $\text{EMCB}\left(\Delta\right)$, however, cannot be attained in an analogous way.

\subsection{Smith Equation \& ARL Expression}\label{ana_arl}
Let $\delta$ denote the ARL of the two targets in our model. In light of Smith's criterion \cite{S05}, these two targets can be resolved w.r.t. their electrical angles if $\Delta$ is \emph{greater} than the standard deviation of the estimate of $\Delta$ (denoted by $\sigma_{\Delta}$) \footnote{Here we assume, without loss of generality, that $\Delta>0$.} . Hence, the ARL $\delta$, being \textit{per definitionem} the lower limit of $\Delta$ that fulfills the above criterion, is identical to the value of $\Delta$ for which $\Delta^2=\sigma_{\Delta}^2$ holds. Furthermore, it is known that under mild conditions \cite{leh83} $\sigma_{\Delta}\approx\sqrt{\mbox{CRB}(\Delta)}$, therefore the value of $\delta$ can computed as the solution to the following equation:
\begin{equation}\label{wc3}
\Delta^{2}=\mbox{CRB}(\Delta),
\end{equation}which is referred to, conventionally, as the \emph{Smith equation}.

The solution of the Smith equation Eq.~(\ref{wc3}) is given by substituting Eqs.~(\ref{f11})-(\ref{f22}) into Eq.~(\ref{14}) and then combining the latter with Eq.~(\ref{wc3}). In doing so, we omit all the terms containing $\Delta^n,\ n>4$, to make the equation easier to solve. Besides, we know from the parameter transformation property of the CRB \cite{Kay93} that $\mbox{CRB}(\Delta)=\mbox{CRB}(-\Delta)$, meaning if $\Delta$ is a root of (\ref{wc3}), then $-\Delta$ will also be a root thereof, thus allowing us to justifiably remove those terms in the equation that contain $\Delta^n,\ n=1,3$ (odd powers of $\Delta$). As a result, we obtain the following quartic equation of $\Delta$:

\begin{equation}\label{12b}
A\Delta^4-B\Delta^2-C=0,
\end{equation}
where
\begin{subequations}
{\setlength\arraycolsep{0.1em}
\begin{eqnarray}
&A=&\frac{2\kappa|\alpha_2|^2}{N}\big(\gamma_{11}\gamma_{22}\gamma_{33}
+2\overline{\gamma_{13}\gamma_{12}^\ast\gamma_{23}^\ast}\nonumber\\
&&-\gamma_{11}|\gamma_{23}|^2-\gamma_{22}|\gamma_{13}|^2
-\gamma_{33}|\gamma_{12}|^2\big),\label{A}\\
&B=&\gamma_{11}\gamma_{33}-|\gamma_{13}|^2,\label{B}\\
&C=&\gamma_{11}\gamma_{22}-|\gamma_{12}|^2;\label{C}
\end{eqnarray}}\end{subequations}in which $(\cdot)^\ast$ denotes the complex conjugate, and $\gamma_{ij},\ i,j=1,2,3$ are defined in Eq.~(\ref{gamma}).

The ARL $\delta$ is taken as the positive real root of Eq.~(\ref{12b}), namely:
\begin{equation}\label{rt}
\delta=\sqrt{\frac{B+\sqrt{B^2+4AC}}{2A}},
\end{equation}while the other roots are trivial and rejected.

\subsection{Existence of the Valid Root}
We remark that Eq.~(\ref{A}) can be reformulated as:
\begin{equation}\label{A2}
A=\frac{2\kappa|\alpha_2|^2}{N}\left|\boldsymbol{\Gamma}\right|,
\end{equation}in which
$\boldsymbol{\Gamma}$ is a $3\times 3$ Gramian matrix whose entries are:
\begin{equation}
\left[\boldsymbol{\Gamma}\right]_{i,j}=\gamma_{ij}=\boldsymbol{\varrho}_i^H\boldsymbol{\varrho}_j,\quad i,j=1,2,3,
\end{equation}where $\boldsymbol{\varrho}_i=\boldsymbol{\Upsilon}^{\frac{1}{2}}\boldsymbol{\rho}_i$ ($\boldsymbol{\Upsilon}$ is defined in Appendix \ref{appendix}.) From Eqs.~(\ref{rho1})-(\ref{rho3}) it is clear that $\boldsymbol{\varrho}_i,\ i,j=1,2,3$, are linearly independent from one another, unless when $\boldsymbol{d}_{\mathcal T}=p\textbf{1}_M$ and $\boldsymbol{d}_{\mathcal R}=q\textbf{1}_N$, where $\textbf{1}_M$ and $\textbf{1}_N$ represent the ones vectors of dimension $M$ and $N$, respectively, $p$ and $q$ are constants not both zero, which occurs only when the inter-sensor spacings at both the transmitter and the receiver all become zero, which is an invalid condition in practice. Thus the Gramian matrix $\boldsymbol{\Gamma}$ is \emph{positive definite}, and $A>0$.

Meanwhile, we can show that $B>0$ and $C>0$ by employing the Cauchy-Schwarz inequality to Eqs.~(\ref{B}) and (\ref{C}); here the equality also holds only under the invalid condition explained above. Now, it follows that $B^2+4AC>0$, signifying that the quadratic equation Eq.~(\ref{12b}) has two distinct real roots, of which our expression in Eq.~(\ref{rt}) is the positive one.

\subsection{Asymptotic expression of $\delta$}\label{alt_arl}
The expression in Eq.~(\ref{rt}) has room for further simplification. Consider the structure of $\gamma_{ij}$ in Eq.~(\ref{gamma}):
\begin{equation}\label{gamma}
\begin{aligned}
\gamma_{ij}&=\boldsymbol{\rho}_{i}^{H}\boldsymbol{\Upsilon}\boldsymbol{\rho}_{j}
=\frac{1}{\sigma^2}\sum_{t=1}^T\boldsymbol{s}^H(t)\boldsymbol{R}_i^H\check{\boldsymbol{\Sigma}}^{-1}
\boldsymbol{R}_j\boldsymbol{s}(t)\\
&=\frac{1}{\sigma^2}\sum_{t=1}^T\sum_{n=1}^N\lambda_n\left[\boldsymbol{U}\boldsymbol{R}_i
\boldsymbol{s}(t)\right]^{H}\left[\boldsymbol{U}\boldsymbol{R}_j
\boldsymbol{s}(t)\right],
\end{aligned}
\end{equation}in which $\boldsymbol{U}$ is a the matrix containing the singular vectors of $\check{\boldsymbol{\Sigma}}^{-1}$, with corresponding eigenvalues denoted as $\lambda_n,\ n=1,\dots,N$. From Eq.~(\ref{gamma}), it is apparent that in the asymptotic cases, e.g., large $T$, $N$, or high SCR (which signifies large $\sum_{t=1}^T\vecnm{\boldsymbol{s}(t)}^2$ or small $\sigma^2$), we have $\gamma_{ij}\gg0$. Furthermore, since from Eqs.~(\ref{A})-(\ref{C}) we have asymptotically that $A=O(\gamma_{ij}^3)$, $B=O(\gamma_{ij}^2)$, and $C=O(\gamma_{ij}^2)$, thus $(B/2A)^2=O(\gamma_{ij}^{-2})\ll(C/A)=O(\gamma_{ij}^{-1})$, which, applied consecutively to Eq.~(\ref{rt}), results in:
\begin{equation}\label{rt2}
\delta=\sqrt{\frac{B}{2A}+\sqrt{\left(\frac{B}{2A}\right)^2+\frac{C}{A}}}\approx\sqrt[4]{\frac{C}{A}},
\end{equation}which is our proposed asymptotic expression for $\delta$.

\subsection{ARL and the Texture Parameters}\label{arl_para}
Eq.~(\ref{rt2}) is not only more concise in form, but allows us to reveal the relationship between the ARL and the texture parameters of the clutter. The derivation follows similar steps as in Subsection \ref{crlb_para}.

First, note that Eq.~(\ref{gamma}) shows $\gamma_{ij}\propto1/\sigma^2$, which, applied to Eqs.~(\ref{A}) and (\ref{C}), leads to $A\propto \kappa/(\sigma^2)^3$ and $C\propto 1/(\sigma^2)^2$. With Eq.~(\ref{rt2}) it then follows that:
\begin{equation}\label{delta_prop}
\delta\propto\sqrt[4]{\sigma^2/\kappa},
\end{equation}and further, by invoking Eqs.~(\ref{a_prop}) and (\ref{kappa_a_prop}), that:
\begin{equation}\label{A_prop}
\delta\propto
\left\{
\begin{aligned}
&\sqrt[4]{\frac{2^{N+a-2}\Gamma(a)}{a\int_0^{+\infty}x^{N+a-1}\frac{K_{a-N-1}^2(x)}
{K_{a-N}(x)}\text{d}x}}, \quad \text{K-distributed clutter}, \\
&\sqrt[4]{\frac{(a+N+1)(a-1)}{a(a+N)}},\quad  \text{t-distributed clutter, for $a>1$}.
\end{aligned}
\right.
\end{equation}In both cases $\delta$ decreases as $a$ increases\footnote{Again, this relationship for a K-distributed clutter can only be determined numerically.}, viz., the ARL is \emph{positively correlated} with $a$.

Furthermore, by combining Eqs.~(\ref{delta_prop}), (\ref{b_prop}) and (\ref{kappa_b_prop}), we observe the independence of the ARL of the scale parameter $b$ under both forms of clutter.

The impact of the texture parameters on the ARL is thus in accordance to that on the CRLBs, and will likewise be certified by our simulation.

\subsection{ARL Based on other CRLBs}\label{relation_arl}
Apart from the ARL based on the standard CRB, one can also obtain its variants based on each of the other CRLBs discussed in Section \ref{sec4}, by equating $\Delta^2$ to the specific CRLB and finding its valid root. For the ARL based on the EMCB, no closed-form expression seems attainable, and its value can be numerically evaluated by the procedure we used in \cite{xzh3}. For the ARL based on the MCRB/HCRB, on the other hand, one can use the analytical expression of $\text{MCRB}\left(\Delta\right)$ or $\text{HCRB}\left(\Delta\right)$ proposed at the end of Subsection \ref{ana_crb} and obtain an analytical expression for $\delta$ by following the same procedure as that in Subsection \ref{ana_arl}. In this case, $\delta$ retains the expression as Eqs.~(\ref{rt}) and (\ref{rt2}), with only the difference that in the expression of $A$ in Eq.~(\ref{A}) $\kappa / N$ is replaced by $\nu$.

\section{Numerical Illustrations}\label{sec6}
In our simulations we consider, unless otherwise stipulated, a collocated MIMO radar comprising $M=5$ sensors at the transmitter and $N=4$ at the receiver, both with half-wave length inter-element spacing. The DOD/DOA of the first target is $60^\circ$, and the angular spacing $\Delta$ between the targets has the value of $1$. Furthermore, the coefficients $\alpha_1$ and $\alpha_2$ are chosen to be $2+0.5j$ and $1-3j$, respectively. The snapshot number $T=6$. Both the real and imaginary parts of the entries of the target source vectors $\boldsymbol{s}(t)$ are generated within the interval $[-1,1]$. For K-distributed clutter, we choose $a=2$ and $b=10$; and for t-distributed clutter, $a=1.1$ and $b=2$. The entries of the speckle covariance matrix $\boldsymbol{\Sigma}$ are generated by $[\boldsymbol{\Sigma}]_{m,n}=\sigma^2\cdot0.9^{|m-n|}e^{j\frac{\pi}{2}(m-n)}, \ m,n=1,\dots,N$ \cite{VSO97}. The SCR is $0$ dB and the number of Monte-Carlo trials is $500$.

\begin{figure}[htpb]
  \centerline{\includegraphics[width=0.5\textwidth]{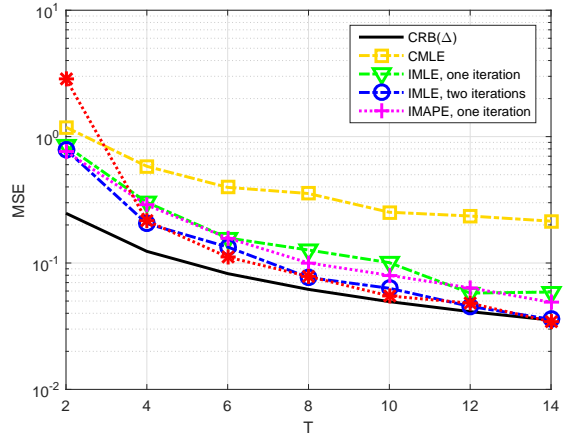}}
  \caption{$\text{MSE}(\Delta)$ vs. $T$ under K-distributed clutter, $\text{SCR}=10$ dB.}
  \label{fig1}
  \end{figure}

\begin{figure}[htpb]
  \centerline{\includegraphics[width=0.5\textwidth]{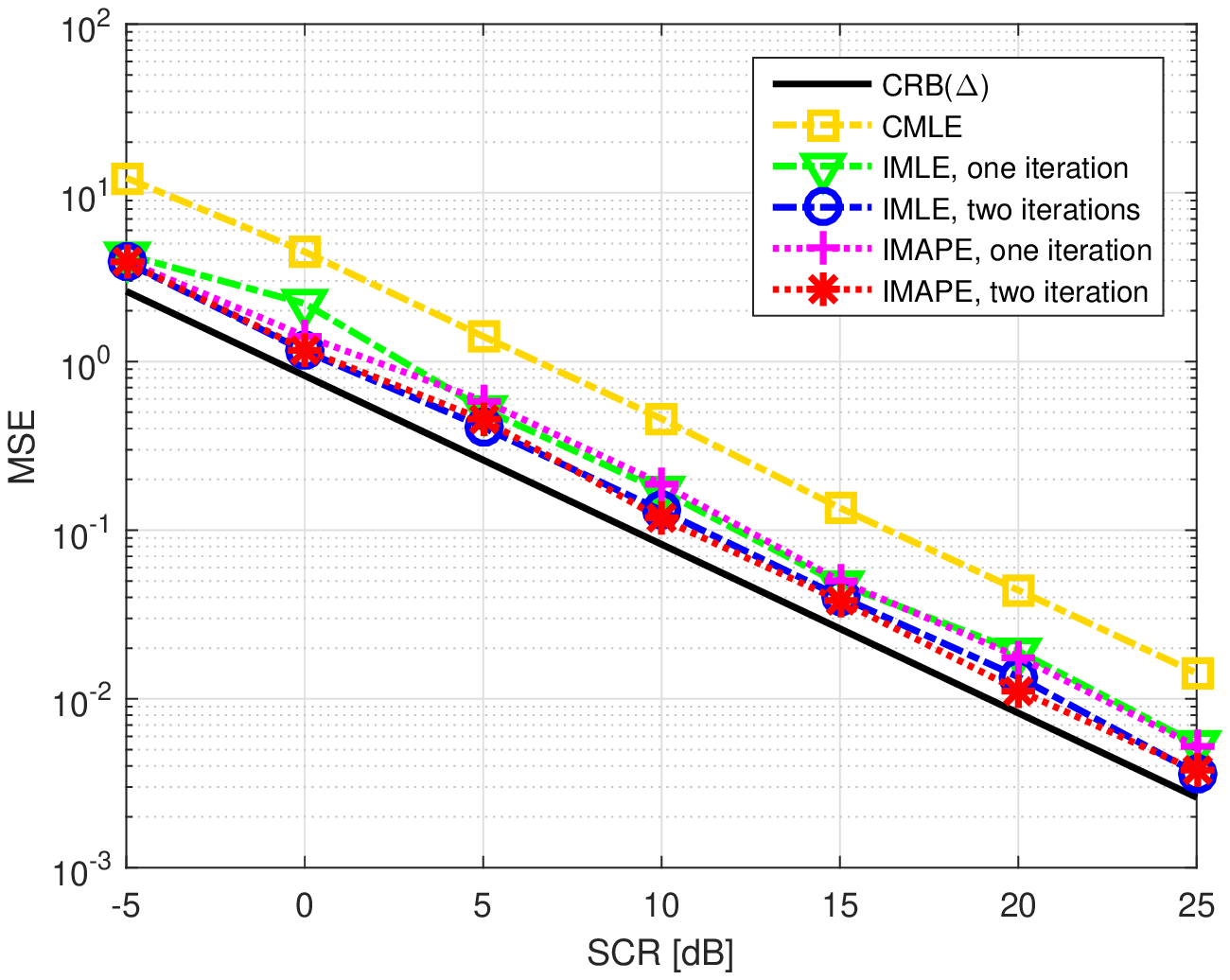}}
\caption{$\text{MSE}(\Delta)$ vs. SCR under K-distributed clutter.}
\label{fig2}
\end{figure}

\begin{figure}[htpb]
  \centerline{\includegraphics[width=0.5\textwidth]{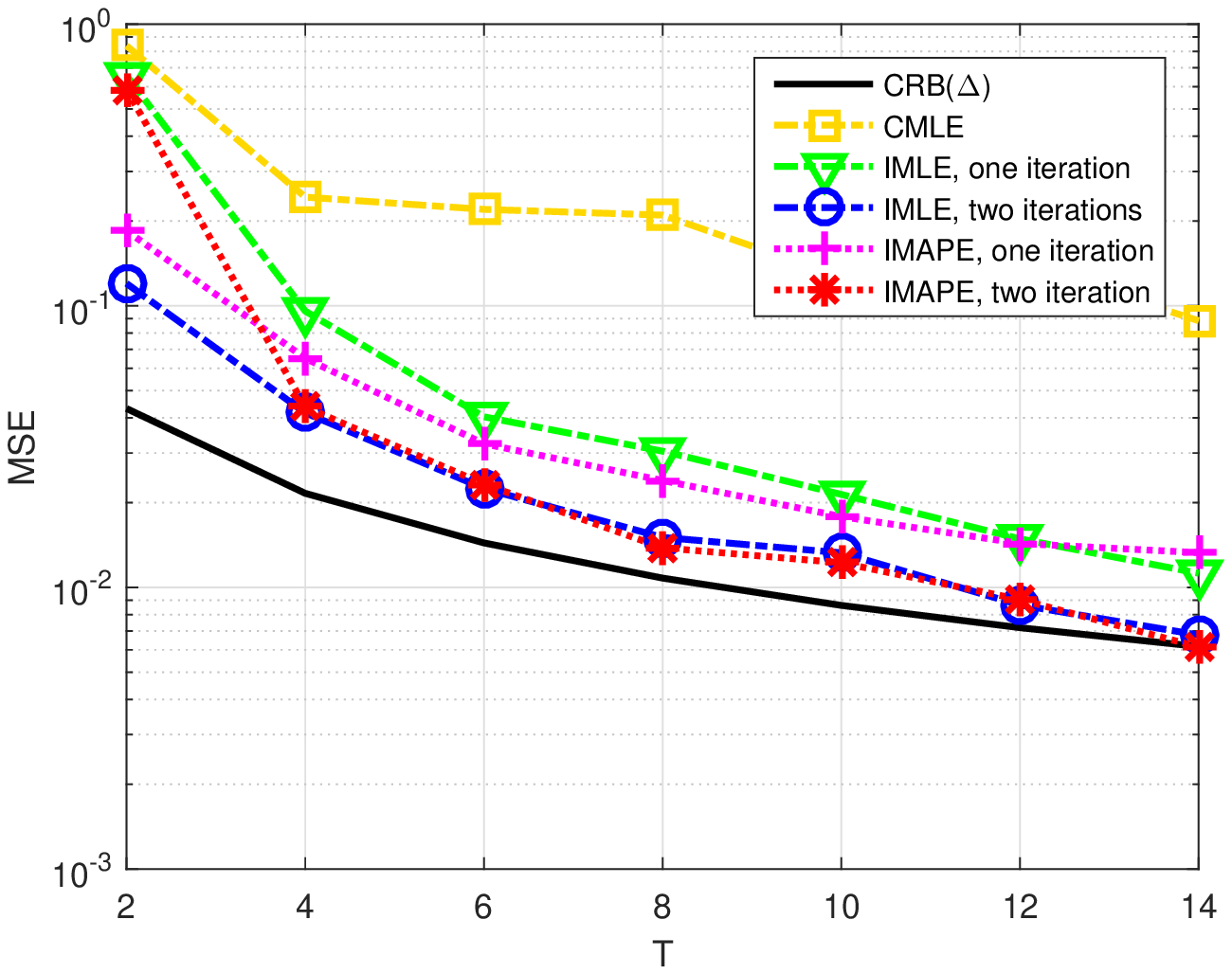}}
  \caption{$\text{MSE}(\Delta)$ vs. $T$ under t-distributed clutter, $\text{SCR}=10$ dB.}
  \label{fig1a}
  \end{figure}

\begin{figure}[htpb]
  \centerline{\includegraphics[width=0.5\textwidth]{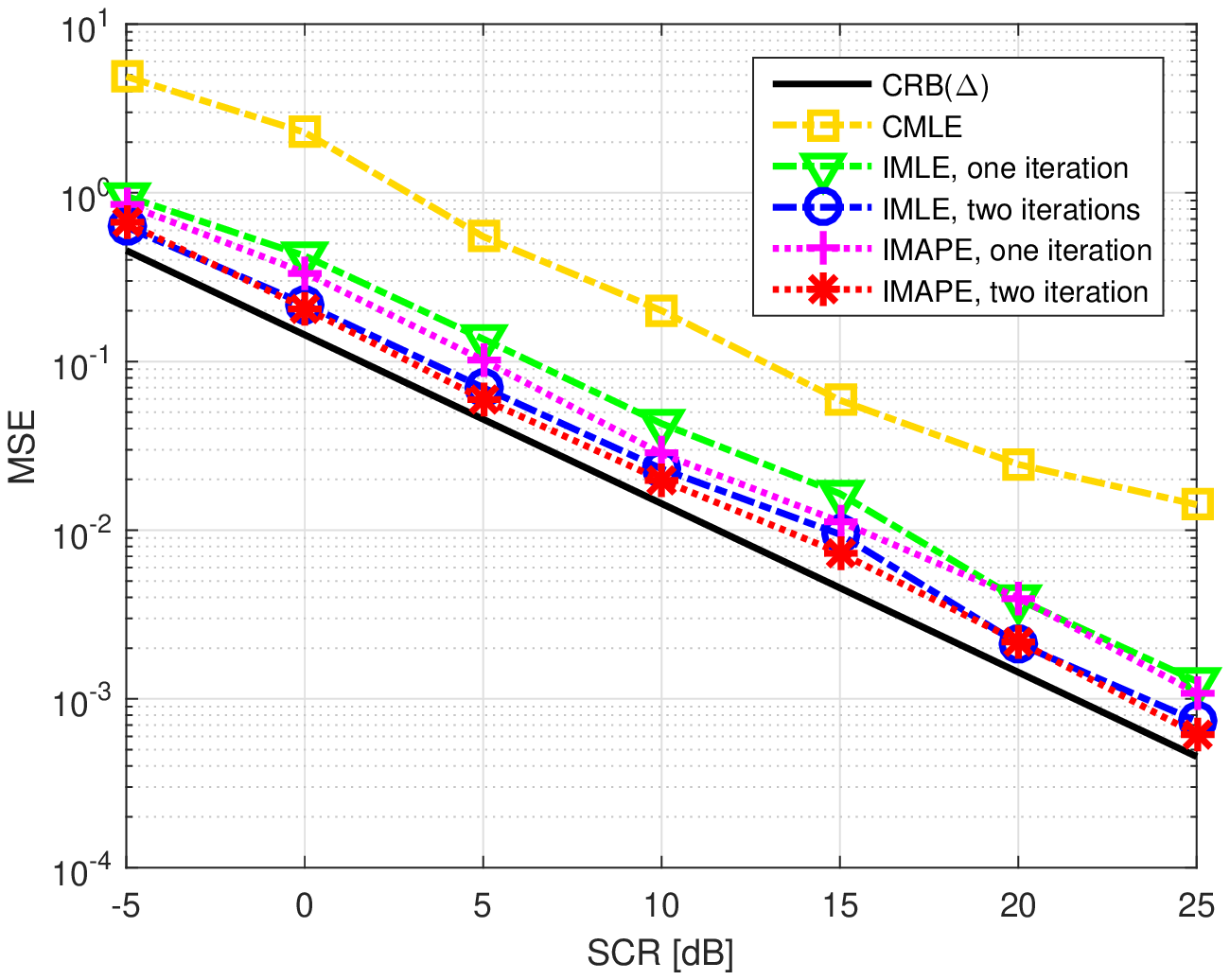}}
\caption{$\text{MSE}(\Delta)$ vs. SCR under t-distributed clutter.}
\label{fig2a}
\end{figure}

In Figs.~\ref{fig1} and \ref{fig2}, we plot the MSEs of the estimation of $\Delta$ under a K-distributed clutter, and in Figs.~\ref{fig1a} and \ref{fig2a} under a t-distributed clutter, versus the snapshot number $T$ and the SCR, respectively. The MSEs are obtained by implementing the CMLE in Eq.~(\ref{conv_ml}) and our proposed IMLE and IMAPE, and are compared with $\text{CRB}(\Delta)$ derived in Subsection \ref{subsec_scrb}. From these four figures, it becomes apparent that the conventional algorithm becomes poor when the clutter is follows a SIRP, and the proposed algorithms lead to far superior performance. The figures also show that as few as two iterations are sufficient for both of our algorithms to have a satisfactory performance in terms of a resulting MSE appropriately close to $\text{CRB}(\Delta)$, in asymptotic $T$ and SCR cases.

\begin{figure}[htpb]
  \centerline{\includegraphics[width=0.5\textwidth]{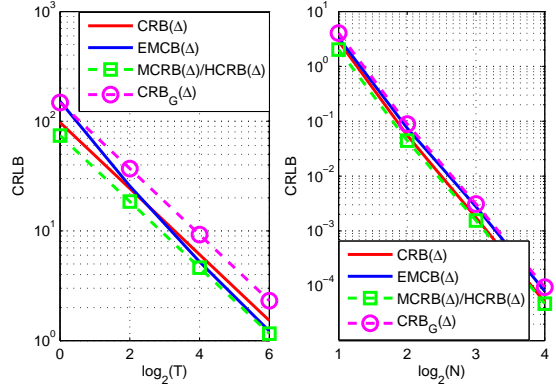}}
  \caption{Left: CRLBs vs. $T$, $M=6$, $N=3$; right: CRLBs vs. $N$, $M=6$, $T=2$. Both under K-distributed clutter.}
  \label{fig3}
  \end{figure}

\begin{figure}[htpb]
  \centerline{\includegraphics[width=0.5\textwidth]{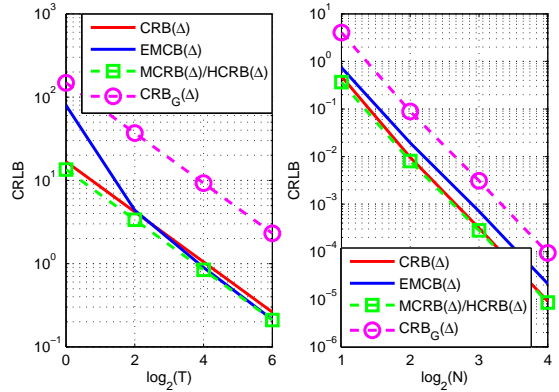}}
  \caption{Left: CRLBs vs. $T$, $M=6$, $N=3$; right: CRLBs vs. $N$, $M=6$, $T=2$. Both under t-distributed clutter.}
  \label{fig3a}
\end{figure}

In Fig.~\ref{fig3}, we plot the CRLBs derived in Section \ref{sec3} under K-distributed clutter, and in Fig.~\ref{fig3a} under t-distributed clutter, versus $T$ and $N$, respectively. In both figures, we add, for comparison, the CRB under Gaussian clutter assumption (denoted by $\text{CRB}_\text{G}(\Delta)$, for which $\kappa=N$). From the figures, we notice that these bounds exhibit exactly the same relationships as were explained in Subsection \ref{relation}, namely, that both the EMCB and the standard CRB is larger than the MCRB/HCRB, to which the EMCB approaches as $T$ gets larger, or the CRB approaches as $N$ does. Furthermore, the EMCB is indifferent to the change of $N$, and the CRB to that of $T$, in terms of their relative distance to the MCRB/HCRB. Which of the two is larger is then \emph{indefinite} and depends on the specific choice of $T$ and $N$. Furthermore, one can see that the CRB under a SIRP clutter assumption is lower than that under the Gaussian one, which is in accordance with the result in \cite{stoica1}, where it was proved that the CRB under the Gaussian data assumption is the \emph{worst-case} one.

\begin{figure}[htpb]
  \centerline{\includegraphics[width=0.5\textwidth]{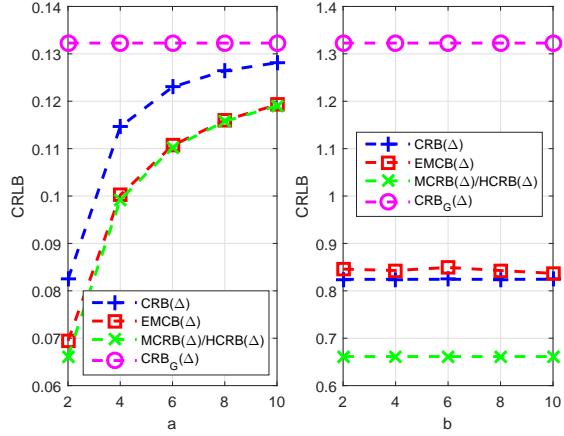}}
  \caption{Left: CRLBs vs. $a$; right: CRLBs vs. $b$. Both under K-distributed clutter.}
  \label{fig3_1}
  \end{figure}

\begin{figure}[htpb]
  \centerline{\includegraphics[width=0.5\textwidth]{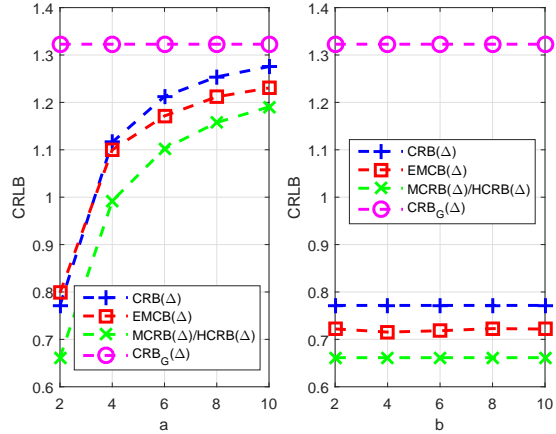}}
  \caption{Left: CRLBs vs. $a$; right: CRLBs vs. $b$. Both under t-distributed clutter.}
  \label{fig3_1a}
\end{figure}

In Fig.~\ref{fig3_1}, we inspect the impact of the texture parameters on the CRLBs under K-distributed clutter, and in Fig.~\ref{fig3_1a} under t-distributed clutter, by plotting, in the left part of both figures, the CRLBs versus $a$ under fixed $b$, and in the right versus $b$ under fixed $a$ ($\text{CRB}_\text{G}(\Delta)$ is also plotted in all the four cases for comparison). The results are in exact accordance with what we have discussed in Subsection \ref{crlb_para}, that for both clutter distributions, the CRLBs increase with $a$ and remain indifferent to the change of $b$. It is notable that the EMCB under K-distributed clutter, whose relationship with $a$ and $b$ has not been analytically established, also follows the same rule as the other CRLBs.

\begin{figure}[htpb]
  \centerline{\includegraphics[width=0.5\textwidth]{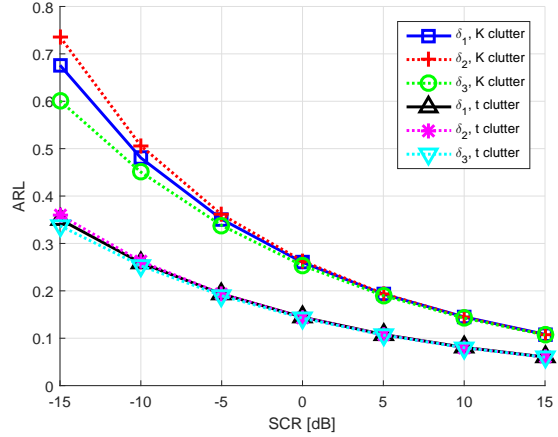}}
  \caption{ARL vs. SCR, M=6, N=8, under K-distributed and t-distributed clutters.}
  \label{fig4}
\end{figure}

In Fig.~\ref{fig4}, we verify, under both K-distributed and t-distributed clutters, our proposed analytical expressions of the ARL in Eqs.~(\ref{rt}) and (\ref{rt2}) (denoted in the figure by $\delta_2$ and $\delta_3$, respectively) by plotting them versus the SCR together with the \emph{exact} ARL (denoted in the figure by $\delta_1$), which is numerically obtained by the approach that we proposed in \cite{xzh3} without any approximation. The figure shows clearly that the values of the three curves essentially coincide in asymptotic cases (above 0 dB in the context) for both distributions of clutter.

\begin{figure}[htpb]
  \centerline{\includegraphics[width=0.5\textwidth]{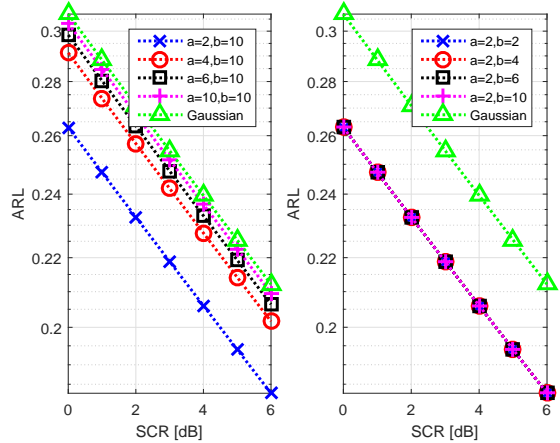}}
  \caption{ARL vs. SCR. Left: varying $a$, fixed $b$; right: varying $b$, fixed $a$. Both with $M=6$ and $N=8$ under K-distributed clutter.}
  \label{fig5}
  \end{figure}
\begin{figure}[htpb]
  \centerline{\includegraphics[width=0.5\textwidth]{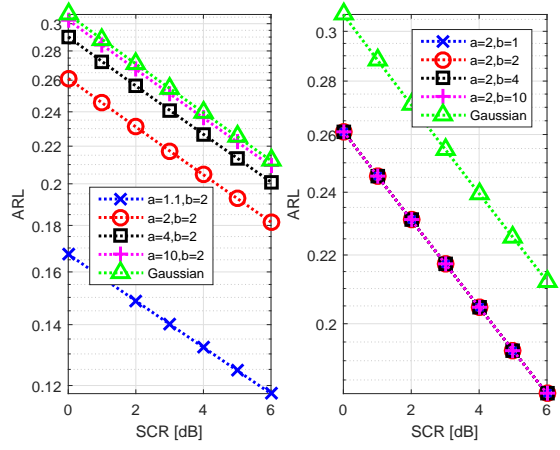}}
  \caption{ARL vs. SCR. Left: varying $a$, fixed $b$; right: varying $b$, fixed $a$. Both with $M=6$ and $N=8$ under t-distributed clutter.}
  \label{fig5a}
\end{figure}

In Figs.~\ref{fig5} and \ref{fig5a}, we investigate the impact of the texture parameters $a$ and $b$ on the ARL under K-distributed and t-distributed clutters, respectively. Again, we fix one of the two parameters and vary the other, and compare the resulting ARLs. One can see from these figures that $\delta$ increases with $a$, but remains invariant w.r.t. changes in $b$, as discussed in Subsection \ref{arl_para}. We also plot the ARL under Gaussian clutter for comparison, which upper-bounds all the ARL results obtained under the various SIRP clutter models considered. In fact we can say, as a direct generalization to the conclusion in \cite{stoica1}, that for given noise power, the targets under Gaussian noise are the most difficult to be correctly resolved.

\begin{figure}[htpb]
  \centerline{\includegraphics[width=0.5\textwidth]{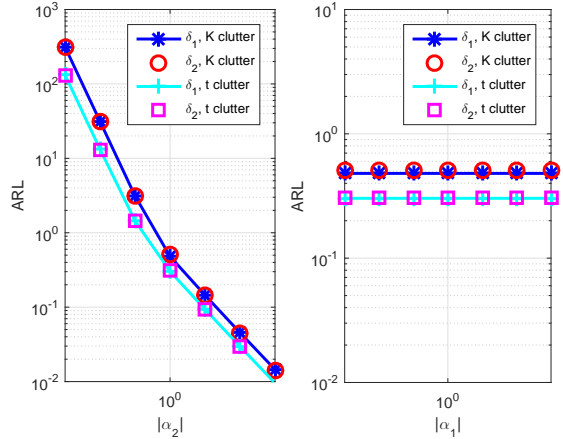}}
  \caption{Left: ARL vs. $|\alpha_2|$, $|\alpha_1|=1$; right: ARL vs. $|\alpha_1|$, $|\alpha_2|=1$. Both with $M=6$ and $N=8$, under K-distributed and t-distributed clutters.}
  \label{fig6}
  \end{figure}

Finally, in Fig.~\ref{fig6} we inspect the respective impact of the power of the two targets on the ARL, by plotting the exact ARL (denoted by $\delta_1$) and the analytical ARL in Eqs.~(\ref{rt}) (denoted by $\delta_2$) for both distributions of clutter, with the power (represented by the absolute value of the RCS factor) of one of the sources fixed and the other varying. From the figure one may observe that, while the ARL decreases with an increasing $|\alpha_2|$, it is independent of the value of $|\alpha_1|$. One may also gain insight into this from our expression in Eq.~(\ref{A}), which is only dependent on $|\alpha_2|$. This follows from the fact that in our model we consider the DOD/DOA of the first source to be known, and the second unknown. Thus, increasing the power of the known source is of no avail in meliorating the resolvability of the sources, and the ARL depends solely on the concrete value of the power of the unknown source, rather than the relative ratio between the power of the two sources.

\section{Conclusion}\label{sec_con}
This paper is dedicated to a systematical investigation into the target estimation and target resolvability problem in a MIMO context under SIRP clutter. We first devised, employing the stepwise numerical concentration approach, two independent but interconnected algorithms, the IMLE and the IMAPE, to deal with the estimation problem of the target spacing. Simulations show that both of our algorithms require only a few iterations to attain convergence, and lead to significantly superior performance than the conventional ML approach.

Next, we derived various CRLBs w.r.t. the target spacing as measures of performance for our algorithms, and analytically compared their relationships. Furthermore, by analytically investigating the effects of the texture parameters on the CRLBs, we found that they all have a positive correlation with the shape parameter, but are all independent of the scale parameter.

We then turned to the resolvability problem, namely, the ARL of two closely-spaced targets. Based on the non-matrix form expression of the CRB w.r.t. the target spacing, which was derived as a by-product, we obtained two analytical expressions for the ARL. We then analyzed the effects of the texture parameters on the ARL, which is analogous to their effect on the CRLBs. Our analytical findings on the CRLBs and the ARL are also numerically corroborated by simulations.

\appendix
\renewcommand{\theequation}{A.\arabic{equation}}
\section{Derivation of the Analytical $\text{CRB}(\Delta)$}\label{appendix}
We follow the same procedure as in Subsection \ref{subsec_scrb} to rederive $\text{CRB}(\Delta)$ based on the model Eq.~(\ref{3}), from which the FIM obtained has exactly the same block structure as shown in Eq.~(\ref{fim_s}). The elements of its parameter block of interest $\boldsymbol{\Phi}'$ have the following expressions:
\begin{subequations}
{\setlength\arraycolsep{0.1em}
\begin{eqnarray}
\phi_{11}'&=&\frac{2\kappa|\alpha_{2}|^{2}}{N}\left(
\gamma_{22}
-4\Delta\widetilde{\gamma}_{23}
+4\Delta^{2}\gamma_{33}
\right)\label{f11}\\
\phi_{22}'&=&\phi_{33}'=\frac{2\kappa}{N}\gamma_{11},\label{f12}\\
\phi_{44}'&=&\phi_{55}'=\frac{2\kappa}{N}\big(
\gamma_{11}
-2\Delta\widetilde{\gamma}_{12}
+\Delta^{2}\gamma_{22}\nonumber\\
&&-2\Delta^{2}\overline{\gamma}_{13}
-2\Delta^{3}\widetilde{\gamma}_{23}
+\Delta^{4}\gamma_{33}
\big),\\
\phi_{12}'&=&\phi_{21}'=\frac{2\kappa}{N}\big(
-\overline{\alpha}_{2}\widetilde{\gamma}_{12}
-\widetilde{\alpha}_{2}\overline{\gamma}_{12}
-2\Delta\overline{\alpha}_{2}\overline{\gamma}_{13}\nonumber\\
&&+2\Delta\widetilde{\alpha}_{2}\widetilde{\gamma}_{13}
\big),\\
\phi_{13}'&=&\phi_{31}'=\frac{2\kappa}{N}\big(
\overline{\alpha}_{2}\overline{\gamma}_{12}
-\widetilde{\alpha}_{2}\widetilde{\gamma}_{12}
-2\Delta\overline{\alpha}_{2}\widetilde{\gamma}_{13}\nonumber\\
&&-2\Delta\widetilde{\alpha}_{2}\overline{\gamma}_{13}
\big),\\
\phi_{14}'&=&\phi_{41}'=\frac{2\kappa}{N}\big(
-\overline{\alpha}_{2}\widetilde{\gamma}_{12}
-\widetilde{\alpha}_{2}\overline{\gamma}_{12}
+\Delta\overline{\alpha}_{2}\gamma_{22}\nonumber\\
&&-2\Delta\overline{\alpha}_{2}\overline{\gamma}_{13}
+2\Delta\widetilde{\alpha}_{2}\widetilde{\gamma}_{13}
-\Delta^{2}\widetilde{\alpha}_{2}\overline{\gamma}_{23}\nonumber\\
&&-3\Delta^{2}\overline{\alpha}_{2}\widetilde{\gamma}_{23}
+2\Delta^{3}\overline{\alpha}_{2}\gamma_{33}
\big),\\
\phi_{15}'&=&\phi_{51}'=\frac{2\kappa}{N}\big(
\overline{\alpha}_{2}\overline{\gamma}_{12}
-\widetilde{\alpha}_{2}\widetilde{\gamma}_{12}
+\Delta\widetilde{\alpha}_{2}\gamma_{22}\nonumber\\
&&-2\Delta\overline{\alpha}_{2}\widetilde{\gamma}_{13}
-2\Delta\widetilde{\alpha}_{2}\overline{\gamma}_{13}
+\Delta^{2}\overline{\alpha}_{2}\overline{\gamma}_{23}\nonumber\\
&&-3\Delta^{2}\widetilde{\alpha}_{2}\widetilde{\gamma}_{23}
+2\Delta^{3}\widetilde{\alpha}_{2}\gamma_{33}
\big),\\
\phi_{23}'&=&\phi_{32}'=\phi_{45}'=\phi_{54}'=0,\\
\phi_{24}'&=&\phi_{42}'=\phi_{35}'=\phi_{53}'\nonumber\\
&=&\frac{2\kappa}{N}\left(
\gamma_{11}
-\Delta\widetilde{\gamma}_{12}-\Delta^{2}\overline{\gamma}_{13}
\right),\\
\phi_{25}'&=&\phi_{52}'=-\phi_{34}'=-\phi_{43}'\nonumber\\
&=&\frac{2\kappa}{N}\left(
-\Delta\overline{\gamma}_{12}+\Delta^{2}\widetilde{\gamma}_{13}
\right)\label{f22},
\end{eqnarray}}\end{subequations}in which
\begin{equation}\label{gamma}
\gamma_{ij}=\boldsymbol{\rho}_{i}^{H}\boldsymbol{\Upsilon}\boldsymbol{\rho}_{j},\quad i,j=1,2,3,
\end{equation}where $\boldsymbol{\rho}_{i}=[\boldsymbol{\rho}_{i}^{T}(1),\dots,\boldsymbol{\rho}_{i}^{T}(T)]^{T}$, $\boldsymbol{\rho}_{j}=[\boldsymbol{\rho}_{j}^{T}(1),\dots,\boldsymbol{\rho}_{j}^{T}(T)]^{T}$, and $\boldsymbol{\Upsilon}=\boldsymbol{I}_{T}\otimes\boldsymbol{\Sigma}^{-1}$. The symbols $\boldsymbol{\Phi}'$ and $\phi_{ij}'$ are used for these to be distinguished from their parallels $\boldsymbol{\Phi}$ and $\phi_{ij}$ derived in Subsection \ref{subsec_scrb} based on the original model.

Let us introduce the following compact block matrix representation of $\boldsymbol{\Phi}'$:
\begin{equation}\label{exp}
\boldsymbol{\boldsymbol{\Phi}'}=\left[
\begin{array}{cc}
\phi_{11}' & \boldsymbol{\varphi}^T \\
\boldsymbol{\varphi} & \boldsymbol{\Omega}
\end{array}\right],
\end{equation}
in which $\boldsymbol{\varphi}=[\phi_{12}',\ \phi_{13}',\ \phi_{14}',\ \phi_{15}']^T$, and
\begin{equation}
\boldsymbol{\Omega}=\left[
\begin{array}{cc}
\boldsymbol{\Omega}_1 & \boldsymbol{\Omega}_2 \\
\boldsymbol{\Omega}_2^T & \boldsymbol{\Omega}_3
\end{array}\right],
\end{equation}where $\boldsymbol{\Omega}_1=\phi_{22}'\boldsymbol{I}_{2}$, $\boldsymbol{\Omega}_3=\phi_{44}'\boldsymbol{I}_{2}$, and
\begin{equation}
\boldsymbol{\Omega}_2=\left[
\begin{array}{cc}
\phi_{24}' & \phi_{25}' \\
-\phi_{25}' & \phi_{24}'
\end{array}\right].
\end{equation}
By employing the block matrix inversion lemma \cite{matrixcookbook} on $\boldsymbol{\Phi}'$ and on $\boldsymbol{\Omega}$ consecutively, we obtain:
\begin{equation}\label{14a}
\text{CRB}\left(\Delta\right)=\left[\boldsymbol{\Phi}'^{-1}\right]_{1,1}
=\frac{\phi_{11}'}{1-\phi_{11}'\boldsymbol{\varphi}^T\boldsymbol{\Omega}^{-1}\boldsymbol{\varphi}},
\end{equation}in which
\begin{equation}\label{bloc}
\boldsymbol{\Omega}^{-1}=\left[
\begin{array}{cc}
\boldsymbol{\Theta}_1& \boldsymbol{\Theta}_2\\
\boldsymbol{\Theta}_3& \boldsymbol{\Theta}_4
\end{array}\right],
\end{equation}where
\begin{subequations}
{\setlength\arraycolsep{0.1em}
\begin{eqnarray}
&\boldsymbol{\Theta}_1=&\left(\boldsymbol{\Omega}_1-\boldsymbol{\Omega}_2\boldsymbol{\Omega}_3^{-1}\boldsymbol{\Omega}_2^T\right)^{-1},\\
&\boldsymbol{\Theta}_2=&-\boldsymbol{\Omega}_1^{-1}\boldsymbol{\Omega}_2\left(\boldsymbol{\Omega}_3
-\boldsymbol{\Omega}_2^T\boldsymbol{\Omega}_1^{-1}\boldsymbol{\Omega}_2\right)^{-1},\\
&\boldsymbol{\Theta}_3=&-\boldsymbol{\Omega}_3^{-1}\boldsymbol{\Omega}_2^T\left(\boldsymbol{\Omega}_1
-\boldsymbol{\Omega}_2\boldsymbol{\Omega}_3^{-1}\boldsymbol{\Omega}_2^T\right)^{-1},\\
&\boldsymbol{\Theta}_4=&\left(\boldsymbol{\Omega}_3-\boldsymbol{\Omega}_2^T\boldsymbol{\Omega}_1^{-1}\boldsymbol{\Omega}_2\right)^{-1};
\end{eqnarray}}\end{subequations}are $2\times 2$ matrices, and $\boldsymbol{\Omega}_1^{-1}$ and $\boldsymbol{\Omega}_3^{-1}$ are simply $1/\phi_{22}'\boldsymbol{I}_{2}$ and $1/\phi_{44}'\boldsymbol{I}_{2}$, respectively.

After calculation, we obtain the analytical expression for $\text{CRB}\left(\Delta\right)$ from Eq.~(\ref{SCRB}) as:
\begin{equation}\label{14a}
\text{CRB}\left(\Delta\right)=\frac{1}{\phi_{11}'+Q},
\end{equation}in which $Q=(\phi_{44}'\phi_{12}'^2+\phi_{44}'\phi_{13}'^2+\phi_{22}'\phi_{14}'^2+\phi_{22}'\phi_{15}'^2
-2\phi_{24}'\phi_{12}'\phi_{14}'-2\phi_{25}'\phi_{12}'\phi_{15}'+2\phi_{25}'\phi_{13}'\phi_{15}'-2\phi_{24}'\phi_{13}'
\phi_{15}')/(\phi_{24}'^2+\phi_{25}'^2-\phi_{22}'\phi_{44}')$.

\section*{References}

\bibliography{zhxtc}

\end{document}